\documentclass[lettersize,journal]{IEEEtran}
\usepackage{amsmath,amsfonts}
\usepackage{algorithmic}
\usepackage{threeparttable}
\usepackage{algorithm}
\usepackage{array}
\usepackage{color}
\usepackage{textcomp}
\usepackage{stfloats}
\usepackage{url}
\usepackage[colorlinks,
          linkcolor = black,		%%修改此处为你想要的颜色
          anchorcolor = black,	%%修改此处为你想要的颜色
          urlcolor = magenta,		%%修改此处为你想要的颜色
          citecolor = black]{hyperref}       % hyperlinks
\usepackage{verbatim}
\usepackage{graphicx}
\usepackage{cite}
\usepackage{svg}
\usepackage{subfigure}
\usepackage[subfigure]{tocloft}
\usepackage{multirow}
\usepackage{makecell}
\usepackage{booktabs}
\usepackage{pifont}       % \ding{xx}
\usepackage{bbding} 
\begin{document}

\title{Enhancing Automatic Modulation Recognition through Robust Global Feature Extraction}
\author{Yunpeng Qu, Zhilin Lu, Rui Zeng,
Jintao Wang,~\IEEEmembership{Senior Member,~IEEE,}

and Jian Wang,~\IEEEmembership{Senior Member,~IEEE}% <-this % stops a space
\thanks{The authors are with the Department of Electronic Engineering, Tsinghua University, Beijing 100084, China (e-mail: qyp21@mails.tsinghua.edu.cn; luzhilin1995@163.com; zeng\_r17@163.com; wangjintao@tsinghua.edu.cn; jian-wang@tsinghua.edu.cn). Corresponding author: Jian Wang.}% <-this % stops a space
%\thanks{Manuscript received April 19, 2021; revised August 16, 2021.}
}

%{Shell \MakeLowercase{\textit{et al.}}: A Sample Article Using IEEEtran.cls %for IEEE Journals}


\maketitle

\begin{abstract}
Automatic Modulation Recognition (AMR) plays a crucial role in wireless communication systems.
Deep learning AMR strategies have achieved tremendous success in recent years.
Modulated signals exhibit long temporal dependencies, and extracting global features is crucial in identifying modulation schemes. 
Traditionally, human experts analyze patterns in constellation diagrams to classify modulation schemes.
Classical convolutional-based networks, due to their limited receptive fields, excel at extracting local features but struggle to capture global relationships.
To address this limitation, we introduce a novel hybrid deep framework named TLDNN, which incorporates the architectures of the transformer and long short-term memory (LSTM). 
We utilize the self-attention mechanism of the transformer to model the global correlations in signal sequences while employing LSTM to enhance the capture of temporal dependencies.
To mitigate the impact like RF fingerprint features and channel characteristics on model generalization, we propose data augmentation strategies known as segment substitution (SS) to enhance the model's robustness to modulation-related features.
Experimental results on widely-used datasets demonstrate that our method achieves state-of-the-art performance and exhibits significant advantages in terms of complexity. Our proposed framework serves as a foundational backbone that can be extended to different datasets. 
We have verified the effectiveness of our augmentation approach in enhancing the generalization of the models, particularly in few-shot scenarios.
Code is available at \url{https://github.com/AMR-Master/TLDNN}.
\end{abstract}


\section{Introduction}
\label{intro}
\IEEEPARstart{T}{he} ability to expeditiously detect and recognize wireless signals has been extensively employed in numerous domains of cognitive radio, such as spectral interference monitoring, dynamic spectrum sensing, communication resource optimization, and a multitude of military applications. The rapid proliferation of mobile devices and the advent of innovative wireless communication technologies have led to a surge in the demand for the utilization of radio frequency (RF) signals \cite{li2019survey}. Therefore, the identification and classification of wireless signals represent a crucial aspect of guaranteeing the stability and efficiency of communication systems \cite{jdid2021machine, zhou2020deep}.

Automatic modulation recognition (AMR) is a pivotal step in wireless signal recognition, where the primary objective is to identify the modulation scheme of received signals while operating under complex environments. 
Traditional AMR approaches can be classified into two categories: likelihood-based (LB) methods and feature-based (FB) methods \cite{dobre2007survey}. 

LB methods employ the likelihood ratio function of the received signal in conjunction with the Bayesian minimum error cost criterion to derive the optimal result \cite{chavali2011maximum,wei2000maximum,7987702}. However, the computational complexity of LB-AMR is exceedingly high, and its implementation necessitates access to a substantial amount of prior information. 
FB methods extract distinctive features from raw data, such as in-phase and quadrature-phase (I/Q) signals, for classification. 
In contrast to the LB-AMR, FB methods are suboptimal from a Bayesian perspective but exhibit reduced complexity. FB techniques leverage various features, including instantaneous time-domain features \cite{nandi1998algorithms}, transform features\cite{lallo1999signal}, spectral features \cite{zhou2017learning}, as well as statistical features such as higher-order cumulants (HOC) \cite{wu2008novel,zhang2018automatic, aslam2012automatic} and higher-order moments (HOM).  
FB methods are dependent on the extraction of a feature set, which requires the manual design of features based on the modulation set and channel characteristics \cite{peng2021survey}.
These methods may no longer be suitable when the communication system and environment change. 

The aforementioned traditional approaches have limitations in their application, highlighting the urgent need to identify a more efficient method for feature extraction from signals.
In recent years, deep learning methods based on neural networks have garnered widespread adoption, achieving state-of-the-art (SOTA) results in the fields of computer vision (CV) \cite{he2016deep} and natural language processing (NLP).
Neural networks are capable of extracting essential hidden information from input data, replacing the manual process of expert feature extraction.
The remarkable achievements have generated significant interest in introducing deep learning to the field of AMR \cite{8054694}.
Convolutional Neural Networks (CNNs), as one of the most widely used network architectures, possess powerful feature extraction capabilities and have been extensively applied in handling RF signal inputs \cite{10080909, 9814482}. O'Shea \emph{et al.} \cite{o2016convolutional} proposed a CNN network and the classic ResNet model for processing both I/Q signal sequences, achieving performance superior to traditional FB methods \cite{o2016convolutional, o2018over}. Zhang \emph{et al.} \cite{9986040} propose high-order convolutional attention networks (HoCANs) to integrate radio signal statistical information with deep networks.

A commonly used alternative approach is to transform the RF signal into an image or matrix \cite{chen2021signet, zeng2019spectrum,peng2018modulation,zha2019deep} like short-time Fourier transform (STFT) \cite{9796039}, thereby converting AMR into a more established image recognition problem. 
For example, Zeng et al. \cite{zeng2019spectrum} applied a Gaussian filter to reduce noise and transformed the signal into a time-frequency map, which was then inputted into a CNN network.
Peng et al. \cite{peng2018modulation} converted the I/Q signals into constellation maps and used CNN-based architectures like AlexNet \cite{krizhevsky2012imagenet} and GoogLeNet \cite{szegedy2015going} for classification.
However, these signal transformation methods lead to an increase in complexity and may result in the loss of critical information during the transformation.

Several studies have employed long short-term memory (LSTM) for extracting time-correlated features in AMR \cite{hong2017automatic}. Rajendran \emph{et al} .\cite{rajendran2018deep} introduced LSTM for processing variable sample lengths and considered network optimization for low-cost devices. 
Ke \emph{et al.} \cite{ke2021real} established a two-layer LSTM autoencoder named LSTM-DAE with a signal reconstruction loss, achieving excellent performance and significantly reducing the complexity.
More recent research has focused on integrating LSTM into CNN-based approaches.
Liu \emph{et al.} \cite{liu2017deep} proposed a CLDNN architecture that consists of two or three convolutional layers followed by two LSTM layers. The CNN layers are proficient in reducing frequency variations, while the LSTM layers are adept at extracting implicit time information.
Xu \emph{et al.} \cite{xu2020spatiotemporal} proposed a multi-channel CNN framework called MCLDNN, which can extract the features from individual and combined I/Q symbols.
Dual-stream\cite{zhang2020automatic}, GrrNet \cite{huang2020automatic}, DCN-BiLSTM \cite{liu2021}, CGDNet \cite{njoku2021cgdnet}, R\&CNN \cite{9694507}, HCGDNN \cite{9764618}, and other networks have adopted similar frameworks that integrate LSTM into the CNN-based architecture.

\begin{figure}[!t]
\centering
\includegraphics[width=0.9\linewidth]{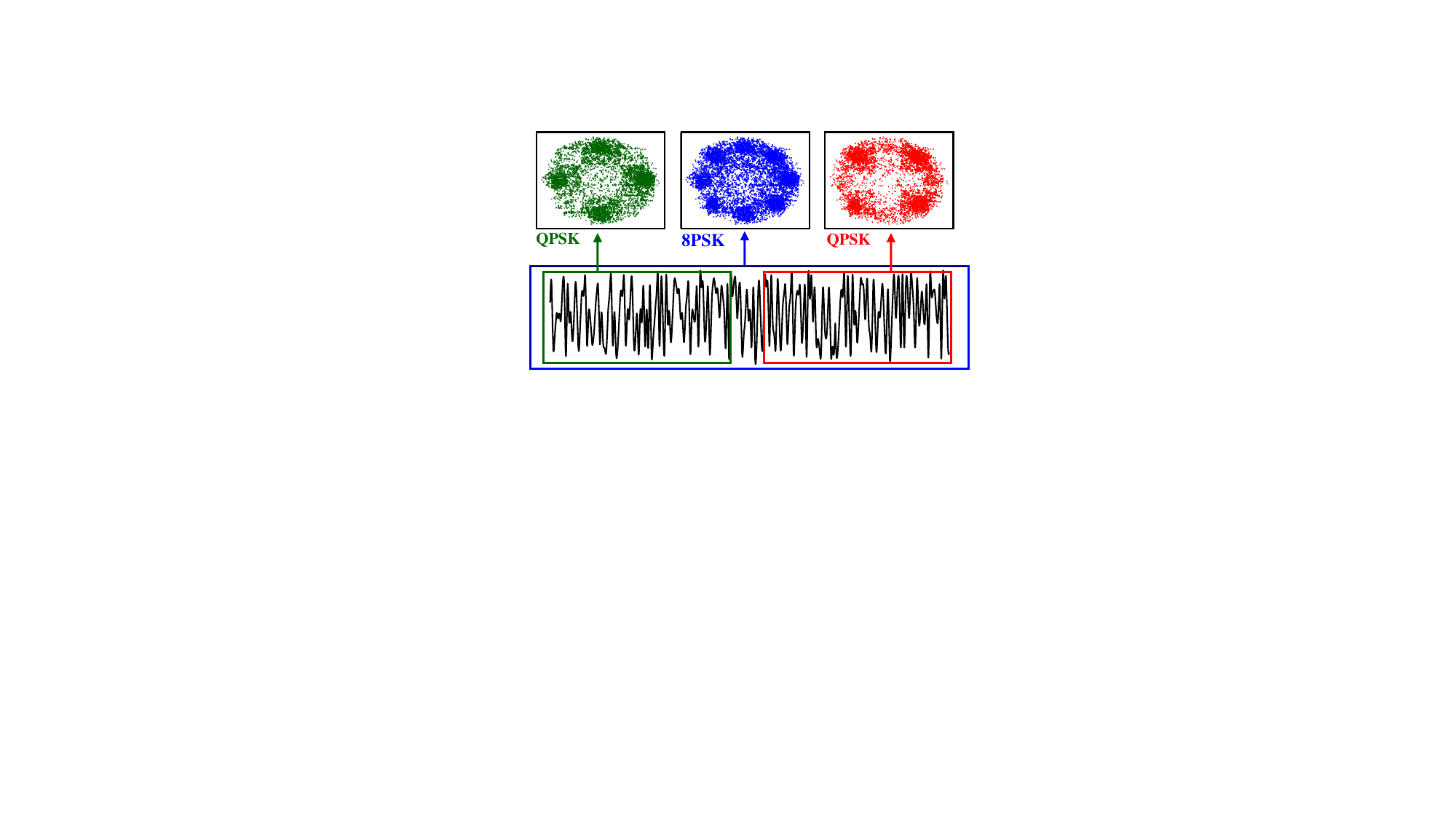}
\caption{Different segments of a signal exhibit different modulation patterns on the constellation diagram.}
\label{constellation}
\end{figure}

Classical AMR methods have often been built upon CNN architectures.
However, the receptive fields of CNNs are typically limited, indicating that they excel at extracting local features rather than global structural features \cite{alzubaidi2021review}.
For AMR, modulated signals possess long-term temporal dependencies. Therefore, we consider the extraction of structural features from a global perspective as the key to identifying the modulation scheme. 
Human experts often classify modulated signals by observing the patterns in the signal's constellation diagram \cite{xiao2022review}.
As the instance shown in Fig. \ref{constellation}, when observing the overall distribution of points in the constellation diagram, it appears to be an 8PSK signal. 
However, upon individual observation of the signal's first half and second half, both constellation diagrams exhibit the pattern of QPSK with or without frequency offset, which is different from the actual 8PSK. 
Similar examples indicate that, when identifying modulation schemes, extracting the global structural features of the signal is more advantageous than focusing on intricate textures. Therefore, an ongoing challenge for AMR lies in devising improved methods to extract these global characteristics.

Another key challenge in AMR is that modulation-related features are often susceptible to interference from other features. For instance, the transmitter may introduce RF fingerprint characteristics due to its inherent imperfections \cite{soltanieh2020review}, while the transmission channel \cite{cavers2005mobile} and the regularity of pilot signals can also introduce irrelevant fixed characteristics.
Obtaining signal samples with labels in real-world scenarios is highly challenging, requiring a more extensive level of specialized expertise and labor compared to other data types like images or audio \cite{chen2021signet}. Therefore, datasets are often collected in similar scenarios and consist of homogeneous data heavily influenced by these inherent features.
In the training process of neural networks, there is a significant risk of overfitting, causing the models to converge towards those irrelevant inherent features and consequently lose their generalization ability to modulation-related characteristics. This issue becomes more prominent when dealing with smaller datasets.

To better capture the global features of signals, we strive to expand the receptive field range through network architecture design.
Transformer \cite{vaswani2017attention} models, which completely discard recursive and convolutional processing by using a multi-head attention mechanism, have an advantage over CNNs. 
Transformer-based models, such as BERT \cite{devlin2018bert}, ViT \cite{50650}, and Swin-T \cite{liu2021swin}, have showcased remarkable performance in various CV and NLP tasks.
Several studies \cite{liu2021automatic}, including MCformer \cite{9685815} and FEA-T \cite{9915584} have attempted to apply transformer to the AMR task and achieve promising results.
However, attention mechanisms do not adequately capture the sequential positional relationships between tokens, which are crucial in processing RF signals, as they contain frequency and phase characteristics. Previous methods have faced challenges in effectively addressing this issue.
LSTM, utilizing gate mechanisms to retain and propagate information, excels in extracting global contextual associations within sequences, thus compensating for the limitations of the transformer.
The combination of LSTM and transformer enables the modeling of global features from two perspectives: attention and temporal relationships. We believe this architecture is effective for the AMR task.
In terms of extracting robust modulation-related features from signals, We draw inspiration from the image domain, where data augmentation is commonly used to expand datasets, aiming to extract generalized features \cite{shorten2019survey}.
We consider analyzing the modulation process of signals and utilizing data augmentation techniques to emphasize modulation-related features that exhibit greater generalization.

In this paper, we propose a hybrid transformer-LSTM deep neural network (TLDNN) backbone, which exploits the self-attention mechanism of the transformer for global information interaction, while utilizing LSTM to extract temporal relationships.
Our framework takes normalized amplitude and phase (A/P) of signals as input and possesses a larger receptive field range compared to conventional CNN-based models, enabling superior extraction of global features.
Furthermore, in the transformer, we incorporate the talking-heads attention mechanism and replace the feed-forward layer (FFN) with rectified gated linear units (ReGLU) to facilitate better interaction within each token embedding.
The proposed framework serves as a backbone network that can be extended to different datasets based on the length or type of signals.
To mitigate the irrelevant inherent impact like RF fingerprint features and channel characteristics, we propose data augmentation strategies known as segment substitution (SS). This method enhances the robustness of the model towards modulation-related features.
The main contributions are as follows.
\begin{enumerate}
\item{
We propose a hybrid neural network backbone based on the transformer and LSTM named TLDNN. 
Our inspiration stems from the research on how human experts utilize constellation diagrams to identify modulation schemes. The crucial aspect of modulation recognition lies in extracting global structural features.
The proposed network, which takes normalized A/P data as input, is capable of extracting global features from two perspectives: attention and temporal relationships.}
\item{
We have conducted explorations on the TLDNN architecture and made two modifications to facilitate information interaction within the transformer: the talking-heads attention and the ReGLU FFN.
Our method achieves significantly superior performance compared to the latest SOTA methods on the widely-used RadioML2016.10a and RadioML2018.01a datasets, particularly in recognizing low signal-to-noise ratio (SNR) signals.
Our method substantially reduces the computational complexity by 80\%-90\% compared to SOTA methods.
Extensive ablation experiments validate the effectiveness of various components and modifications while shedding light on the impact of model depth variations.
}
\item{To mitigate the irrelevant factors like RF fingerprint features and channel characteristics, we propose data augmentation strategies known as segment substitution, where segments of the input signal are randomly replaced.
Our augmentation approach demonstrates its efficacy in improving the generalization and robustness of different baselines.
In few-shot scenarios, modulation features are highly susceptible to interference from those irrelevant inherent features. Our SS strategy achieves a performance improvement of 5\% by effectively enhancing the model's ability to focus on modulation-related features. This is particularly significant in real-world scenarios where obtaining labeled signals is challenging.
}
\end{enumerate}

\section{System Model}
\begin{figure*}[!t]
\centering
\includegraphics[width=0.95\linewidth]{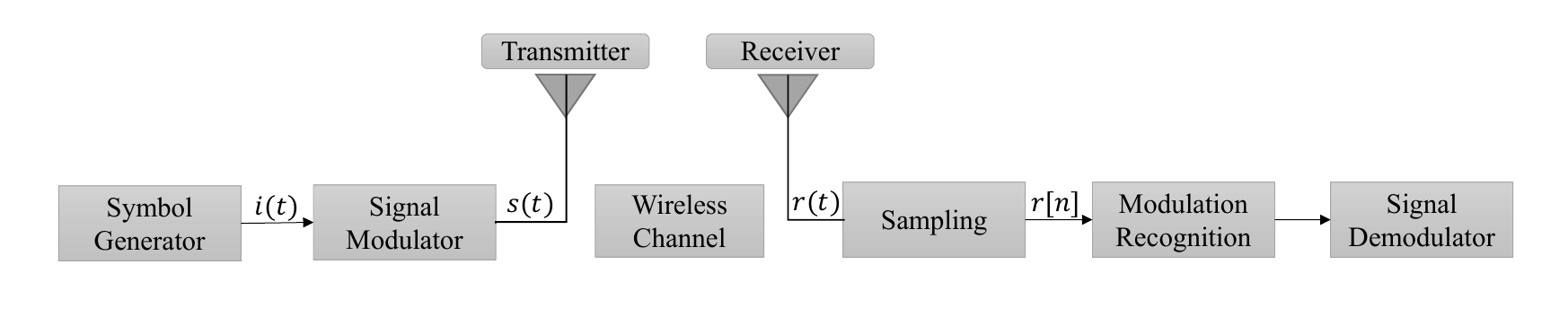}
\caption{Communication system model.}
\label{system}
\end{figure*}
The current general communication system model can be represented in the form illustrated in Fig. \ref{system}. The stream of symbols generated by the source can be described as $i(t)$. The symbol sequence $i(t)$ is modulated to a specific scheme and sent via a transmitting antenna. The signals $s(t)$ emitted by the transmitter can be represented as follows:
\begin{equation}
\label{emitted signal}
s(t) = S(i(t);\boldsymbol{\theta})
\end{equation}
where $S$ represents the function of the modulation process, and $\boldsymbol{\theta}$ represents the set of modulation parameters.

The transmitted signal undergoes propagation through the channel and eventually reaches the receiving antenna. 
Wireless channels encompass various impairments, including but not limited to selective fading, propagation delay, and thermal noise.
The received signal $r(t)$ can be represented as follows:
\begin{equation}
\label{received signal}
r(t) = s(t)*h(t,\tau)+n(t)
\end{equation}
where $h(t,\tau)$, $\tau$, and $n(t)$ represent the channel impulse response function, channel parameters, and the additive Gaussian noise, respectively.
In practice, the continuous signal $r(t)$ is sampled with a fixed sampling rate $f_s$ and subjected to Hilbert transformation when received. The representation of the received time-domain discrete complex signal $r[n]$ is as:
\begin{equation}
\label{descrete signal}
r[n] = r_I[n]+jr_Q[n]
\end{equation}
where the real and imaginary parts $r_I[n]$ and $r_Q[n]$ represent the n-th I/Q component of the signal $r[n]$ with a sequence length of $N$. The signal $r[n]$ can be further represented:
\begin{equation}
\label{amplitude and phase}
r[n] = r_A[n]e^{j\pi r_P[n]}
\end{equation}
where $r_A[n]$ and $r_P[n]$ represent the n-th amplitude and phase component of the signal $r[n]$. 
At the receiver end, the discrete signal $r[n]$ is recognized by the modulation recognition module to determine the modulation type, followed by demodulation to retrieve the original information.

In this article, the signal is transformed into normalized A/P components to standardize the input data.
$r_P[n]$ has been normalized to the range of -1 and 1, with the units being in radians. Such normalization confers advantages in learning temporal dependencies \cite{rajendran2018deep}. 
We adopt the min-max normalization to transfer signal $r_A[n]$ to the range of 0 and 1:
\begin{equation}
\hat{r}_A[n]=\frac{r_A[n]-\min \{\boldsymbol{r_A}\}}{\max \{\boldsymbol{r_A}\} - \min \{\boldsymbol{r_A}\} }
\end{equation}

Neural networks typically take input in the form of vectors or matrices. 
The signal is represented as matrix $\boldsymbol{X} \in \mathbb R^{N\times 2}$ to serve as input for the modulation recognition network.
\begin{equation}
\boldsymbol{X} = 
\begin{bmatrix}
\hat{r}_A[0] & \hat{r}_A[1] & ... & \hat{r}_A[N-1] & \hat{r}_A[N] \\
 r_P[0] &  r_P[1] & ... & r_P[N-1] & r_P[N] \end{bmatrix}^T
\end{equation}

\section{Methods}
\begin{figure*}[!t]
\centering
\includegraphics[width=0.94\linewidth]{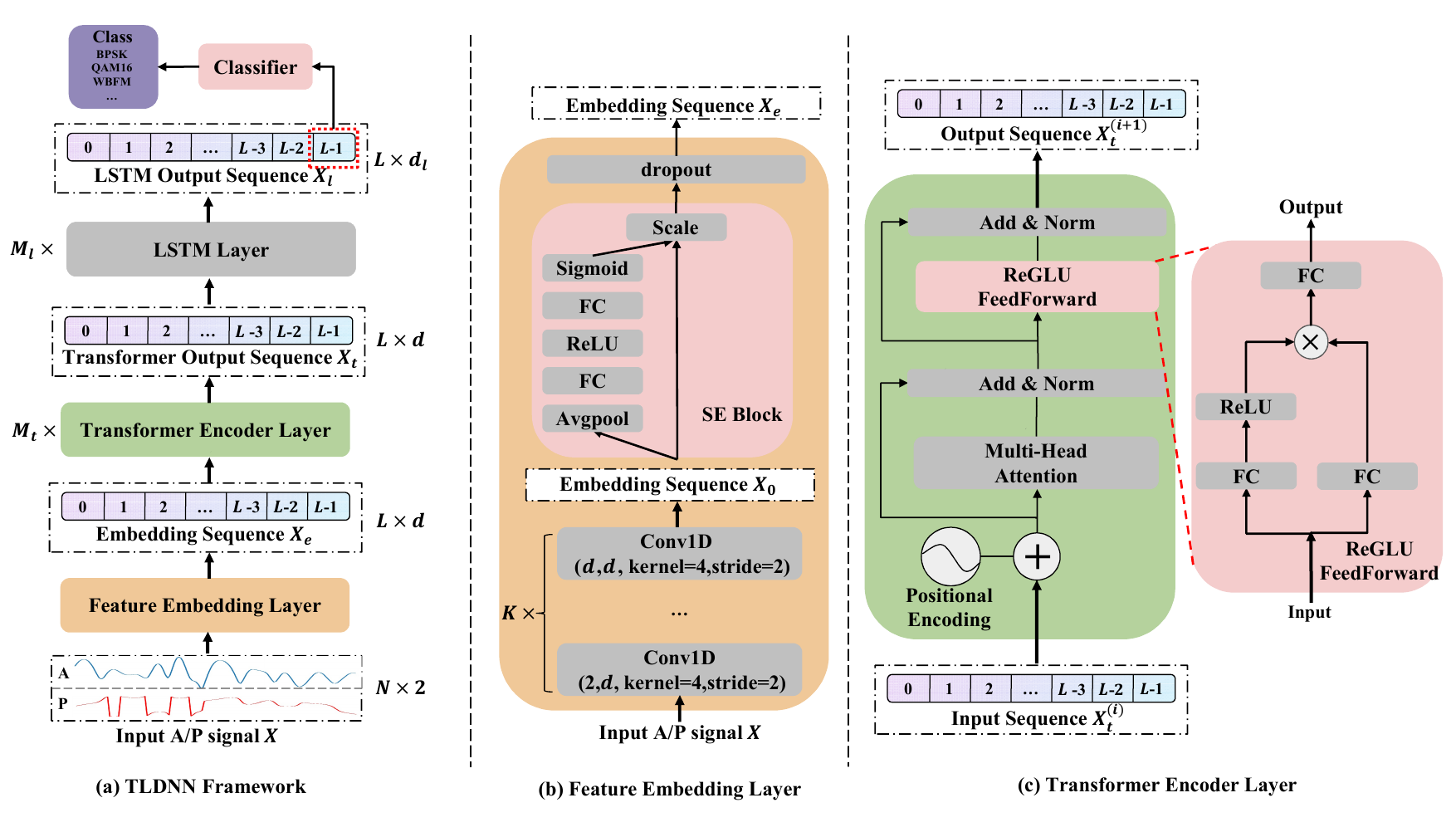}
\caption{The diagram of our proposed architecture.}
\label{model}
\end{figure*}
In this section, we introduce our proposed TLDNN framework and segment substitution strategy.
The diagram of our TLDNN is shown in Fig. \ref{model} (a), where the network takes the normalized A/P signals as input.
We introduce the three components of the overall architecture separately: the feature embedding laye (Sec. \ref{embedding}), the transformer encoder layer (Sec. \ref{tranformerencoder}), and the LSTM layer (Sec. \ref{lstm}). 
The transformer layer, illustrated in Fig. \ref{lstmtransformer} (a), achieves interaction among different tokens through the self-attention mechanism. 
The LSTM layer, as depicted in Fig. \ref{lstmtransformer} (b), facilitates the flow of sequential information. These components enable us to model global information effectively from two perspectives. Detailed explanations will be provided in the subsequent sections.
In Sec. \ref{section:randommixing}, We introduce the motivation and detailed implementation of our proposed SS method for data augmentation.

\subsection{Feature Embedding Layer}
\label{embedding}
\subsubsection{Convolutional Layer}
In our preliminary stage, we design a feature embedding layer based on CNNs to extract localized features from the input signal and map them to token embeddings. 
These convolutional layers can be regarded as filters that map the signal to different frequency domains.
As shown in Fig. \ref{model} (b), the input signals are subjected to $K$ layers of analogous convolutional layers. Within the first convolutional layer, the two-channel vectors on each time sequence are mapped to a $d$-dimensional embedding. 
The subsequent layers perform mappings from $d$-dimensional to $d$-dimensional space to facilitate further feature extraction.
Since we aim to utilize CNNs for extracting local features, the analogous kernels should have smaller local receptive fields. 
The size of the kernels $K_c$ should be smaller than the length of the symbol sequences $L_s$ to avoid extracting features that span across symbols. 
%In the experiments, we have chosen the RadioML dataset series for testing, where each symbol has a sample size of $8$. Therefore,  it is appropriate to set the kernel size to $4$. 
Furthermore, we modify the stride of the convolutional layers to $2$, gradually compressing the sequence length to further enhance the complexity and efficiency. 
Finally, we obtain the feature embedding tokens $\boldsymbol{X}_0\in\mathbb R^{L\times d}$, where each token consists of a $d$-dimensional feature from a sliding receptive field.
The relationship $L=\lfloor \frac{N}{2^K}\rfloor$ signifies that we can adapt the number of layers $K$ based on the lengths of input signals to maintain a stable sequence length when transmitting them to subsequent networks.

\subsubsection{SE Block}
We introduce the Squeeze-and-Excitation (SE) block to adaptively recalibrate channel-wise feature responses by modeling interdependencies between channels \cite{Hu_2018_CVPR}. The global spatial information is squeezed into a channel descriptor 
$\boldsymbol{z}\in\mathbb R^d$ and fed into a formed bottleneck to capture channel-wise dependencies as the weights of each channel $\boldsymbol{s}\in\mathbb R^d$:
\begin{equation}
\label{se1}
\boldsymbol{s}= \sigma(\boldsymbol{W}_2\delta(\boldsymbol{W}_1\boldsymbol{z}))
\end{equation}
where $\delta$ refers to the ReLU function, $\boldsymbol{W}_1\in\mathbb R^{\frac{d}{r}\times d}$ and $\boldsymbol{W}_2\in\mathbb R^{d\times\frac{d}{r}}$. The $i$-th element of $\boldsymbol{z}$ is calculated by:
\begin{equation}
\label{se2}
z_i = \frac{1}{N}\sum_{j=0}^N\boldsymbol{X}_e(i,j)
\end{equation}
The final embedding tokens are obtained by rescaling $\boldsymbol{X_0}$ with the channel weight vector $\boldsymbol{s}$:
\begin{equation}
\label{se3}
\boldsymbol{X}_e = F_{scale}(\boldsymbol{s},\boldsymbol{X}_0)=\boldsymbol{sX}_0
\end{equation}
The presence of a dropout layer helps alleviate the issue of overfitting.

\subsection{Transformer Encoder Layer}
\label{tranformerencoder}
\begin{figure}[!t]
\begin{center}  
    \subfigure[Self-attention mechanism facilitates the global interaction among tokens.]{  \includegraphics[width=0.9\linewidth]{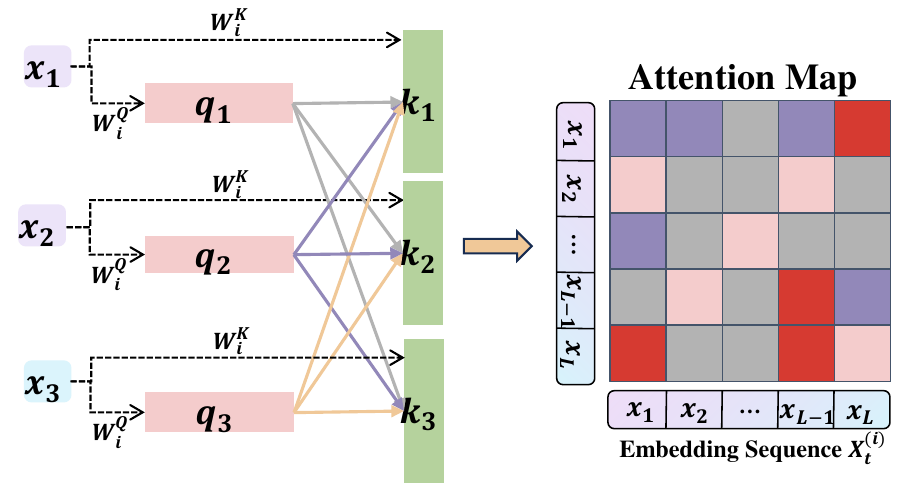}  }  
    \subfigure[LSTM enables the seamless flow of information among tokens.]{  \includegraphics[width=0.9\linewidth]{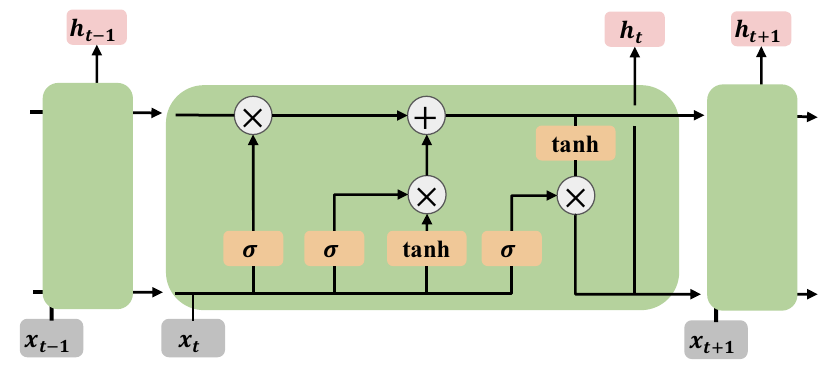}  }  
\caption{The mechanisms of transformer and LSTM for extracting global features.}
\label{lstmtransformer}
\end{center}
\end{figure}
Afterward, the feature embeddings are utilized to extract spatial correlations through $M_t$ identical layers of transformer encoder, which is shown in Fig. \ref{model} (c). In each layer of the transformer encoder, there is a multi-head attention block and a position-wise feed-forward network. We take the initial input $\boldsymbol{X}_e$ as $\boldsymbol{X}_t^{(0)}$ and the final output $\boldsymbol{X}_t$ as $\boldsymbol{X}_t^{(M_t)}$.

\subsubsection{Positional encoding}
Due to the absence of recurrence or convolution in the transformer architecture, the inherent temporal order of sequences cannot be effectively captured. Thus, the incorporation of positional encodings becomes essential to encode and convey the relative or absolute positional information of words within the sequence. 
In this paper, positional encodings are treated as learnable variables with the same dimensions as the input sequence. Consequently, they can be added to the input embeddings.

\subsubsection{Multi-head Attention}
The attention mechanism can perceive the overall structural information by facilitating global interaction among token sequences.
Multi-head attention gives multiple representation subspaces for the attention layer and extends the model’s ability to focus on different locations. 

In the $n$-th layer, the input tokens $\boldsymbol{X}_t^{(n)}$ are multiplied by the linear transformation to get queries, keys, and values of each head. In the $i$-th head among all $h$ heads, they can be represented as $\boldsymbol{Q_i}\in\mathbb R^{L\times d_t}$, $\boldsymbol{K_i}\in\mathbb R^{L\times d_t}$ and $\boldsymbol{V_i}\in\mathbb R^{L\times d_t}$:
\begin{equation}
\label{QKV}
\boldsymbol{Q}_i=\boldsymbol{X}_t^{(n)}\boldsymbol{W}_i^Q,\; \boldsymbol{K}_i=\boldsymbol{X}_t^{(n)}\boldsymbol{W}_i^K,\;
\boldsymbol{V}_i=\boldsymbol{X}_t^{(n)}\boldsymbol{W}_i^V
\end{equation}
where we fix $d_t=\frac{d}{h}$ and $\boldsymbol{W}_{i}^{Q}\in\mathbb{R}^{d\times d_t}, \boldsymbol{W}_{i}^{K}\in\mathbb{R}^{d\times d_t}, \boldsymbol{W}_{i}^{V} \in \mathbb{R}^{d \times d_t}$.
In each head, we compute the dot products of each query with all keys, divide each by $\sqrt{d_t}$, and apply the softmax function to calculate the attention map.
As depicted in Fig. \ref{lstmtransformer} (a), each token is linearly mapped to generate a query and key, which are used for interaction with other tokens. The attention map illustrates the correlation between different tokens in the entire sequence, representing the global structural information.

The attention map is subsequently utilized as similarity weights to multiply with the global sequence value $\boldsymbol{V}_i$, thereby generating a reorganized feature mapping.
In our method, we adopt a variant of multi-head attention called talking-heads attention, which introduces a linear mapping across head dimensions \cite{shazeer2020talkingheads, 9915584}. It brings improvements in perplexity for masked sequence modeling tasks.
The talking-heads mechanism includes linear projections across the attention-heads dimension, immediately before and after the softmax function. The output of the $i$-th head is:
\begin{equation}
\label{attention}
\boldsymbol{head}_i = softmax(\frac{ \boldsymbol{Q}_i (\boldsymbol{K}_i)^T \boldsymbol{P}_l} {\sqrt{d_k}}) \boldsymbol{P}_w\boldsymbol{V}_i
\end{equation}
where talking-heads projections $\boldsymbol{P}_l \in \mathbb{R}^{h\times h}, \boldsymbol{P}_w\in\mathbb{R}^{h \times h}$.
The final result is a concatenation of these heads: 
\begin{equation}
\label{multi-head}
Attention(\boldsymbol{X}_t^{(n)}) = Concat(\boldsymbol{head}_1, ..., \boldsymbol{head}_h)\boldsymbol{W}^O
\end{equation}
where  $\boldsymbol{W}^O\in\mathbb R^{d\times d}$.

\subsubsection{Feed-Forward Network}
Following the attention operation, the output is individually and uniformly processed by a fully connected Feed-Forward Network at each position. comprising two linear transformations and a ReLU activation.
In the original transformer model, the feed-forward layer consists of two linear layers and a ReLU activation function.
Following \cite{shazeer2020glu}, we have substituted the feed-forward network with Rectified Gated Linear Units (ReGLU). By incorporating two parallel branches with gated controls, we aim to enhance the representational capability by enabling interaction between each token vector. The output can be described as:
\begin{equation}
\label{ReGLU}
ReGLU(\boldsymbol{X}_t^{(n)}) =  (\delta(\boldsymbol{X}_t^{(n)}\boldsymbol{W}_1)\otimes (\boldsymbol{X}_t^{(n)}\boldsymbol{W}_2))\boldsymbol{W}_3
\end{equation}
where $\delta$ refers to the ReLU function and  $\otimes$ refers to the element-wise product.

\subsection{LSTM Layer and Classifier}
\label{lstm}
Although positional encodings are used to represent the relative position information of a sequence, the transformer still has limitations in capturing sequence order information, especially in cases where there are strong correlations between the timing of RF signals.
Therefore, the output sequence from transformer encoder layers is further processed by $M_l$ LSTM layers with the hidden size of $d_l$ to extract more profound temporal correlations.
As depicted in Fig. \ref{lstmtransformer} (b), LSTM leverages gate mechanisms to facilitate the storage and flow of information between different tokens in the sequence. 
This allows the LSTM to generate temporal features that incorporate global sequence information, providing a distinct perspective compared to the attention mechanism.

In the end, we utilize the final vector from the LSTM output sequence $\boldsymbol{X}_{l}$ as the input to the classifier, resulting in the modulation recognition types. The classifier is composed of three fully connected layers with ReLU used as the activation function.

\subsection{Segment Substitution}
\label{secton:randommixing}
\begin{figure}[!t]
\begin{center}  
    \subfigure[Signals with the same modulation scheme appear similar in the constellation diagram.]{  \includegraphics[width=0.9\linewidth]{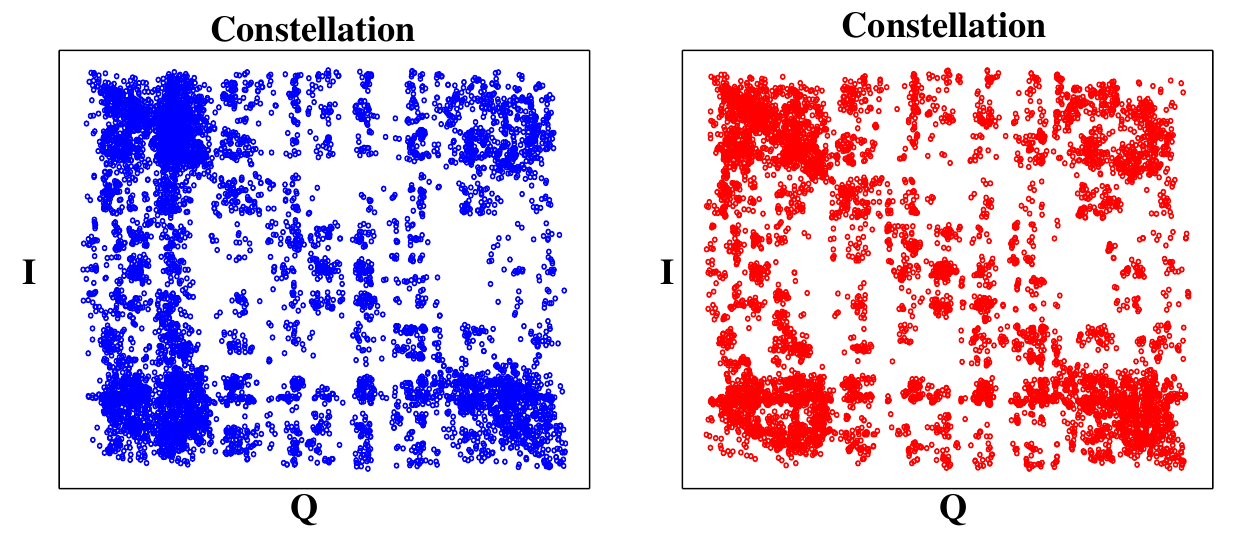}  }  
    \subfigure[Signals with the same modulation scheme can transmit different symbol sequences.]{  \includegraphics[width=0.9\linewidth]{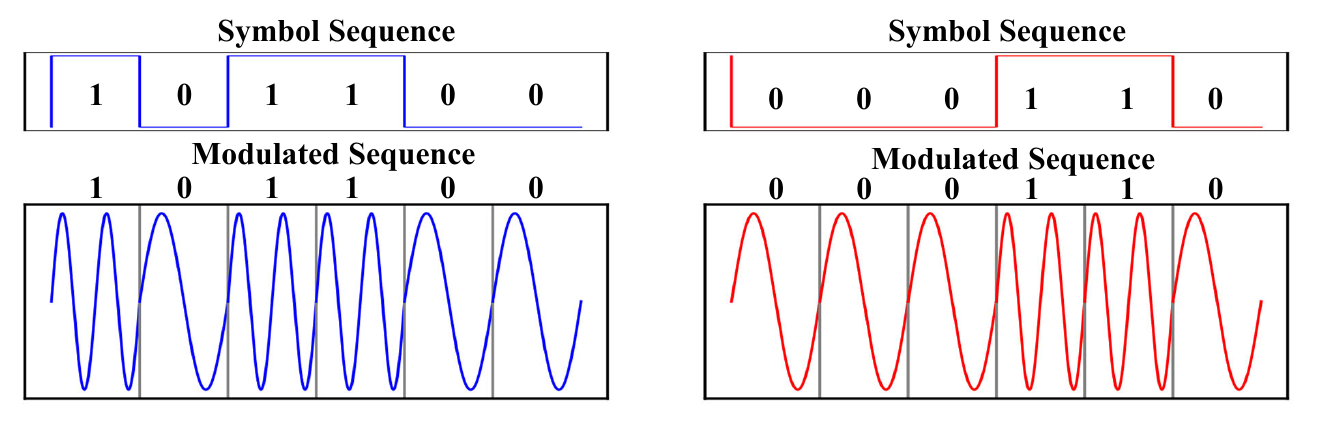}  }  
\caption{The properties of modulation in constellation diagrams and transmission symbols}
\label{constellationsymbol}
\end{center}
\end{figure}
To enhance the model’s robustness and generalization to modulation-related features and minimize interference from irrelevant inherent features such as RF fingerprint characteristics and channel characteristics,  we propose a data augmentation method known as segment substitution.
We analyze the properties of the modulation scheme itself and aim to enhance these properties through data augmentation techniques. Additionally, we intend to disrupt other interfering features that do not exhibit these properties.
Our motivations and methods are described below.

As we analyze in Sec. \ref{intro}, human experts often rely on the constellation diagram to make judgments about the modulation scheme. 
Different modulation schemes typically display distinct patterns on the constellation diagram,  whereas signals with the same modulation scheme exhibit the same constellation pattern.
As shown in Fig. \ref{constellationsymbol} (a), the two signals with the same modulation scheme display similar patterns, with the constellation points clustering around the four corners of a square. Hence, randomly selecting discrete samples within the signal and replacing them with segments of the same modulation scheme would not exhibit distinctions on the constellation diagram. 
Applying this substitution based on the invariance can contribute to enhancing the network's attention to the global distribution of constellation points.
Moreover, since channel models and RF fingerprint features often involve frequency and phase offsets, substituting sample points can be meaningful in disrupting such temporal relationships.
This disruption permits the network to focus less on local information within symbols, thereby facilitating the extraction of more generalized global features.

Another approach revolves around the notion that a sequence of symbols is modulated and mapped to a signal at the transmitter, where each symbol corresponds to several consecutive sample points that are received.
Symbols are independent of each other, and there is no correlation between the choice of symbols and the modulation scheme.
As depicted in Fig. \ref{constellationsymbol} (b), signals with the same modulation scheme can transmit entirely different symbol sequences.
This means that if we replace several symbols in a signal with symbols of the same modulation scheme, we can generate a new sample of that modulation scheme.
Based on this concept, the expansion of the dataset can be achieved by selectively replacing a subset of consecutive sample points within the signal. 
We believe that this substitution, based on the independence of symbols, can also help accentuate the characteristics of the modulation scheme.
Additionally, since the symbols in a pilot signal exhibit strong regularity, this substitution also aids in disrupting the correlation among the pilot symbols.

\begin{figure}[!t]
\begin{center}  
    \subfigure[Discrete Segment Substitution]{  \includegraphics[width=0.95\linewidth]{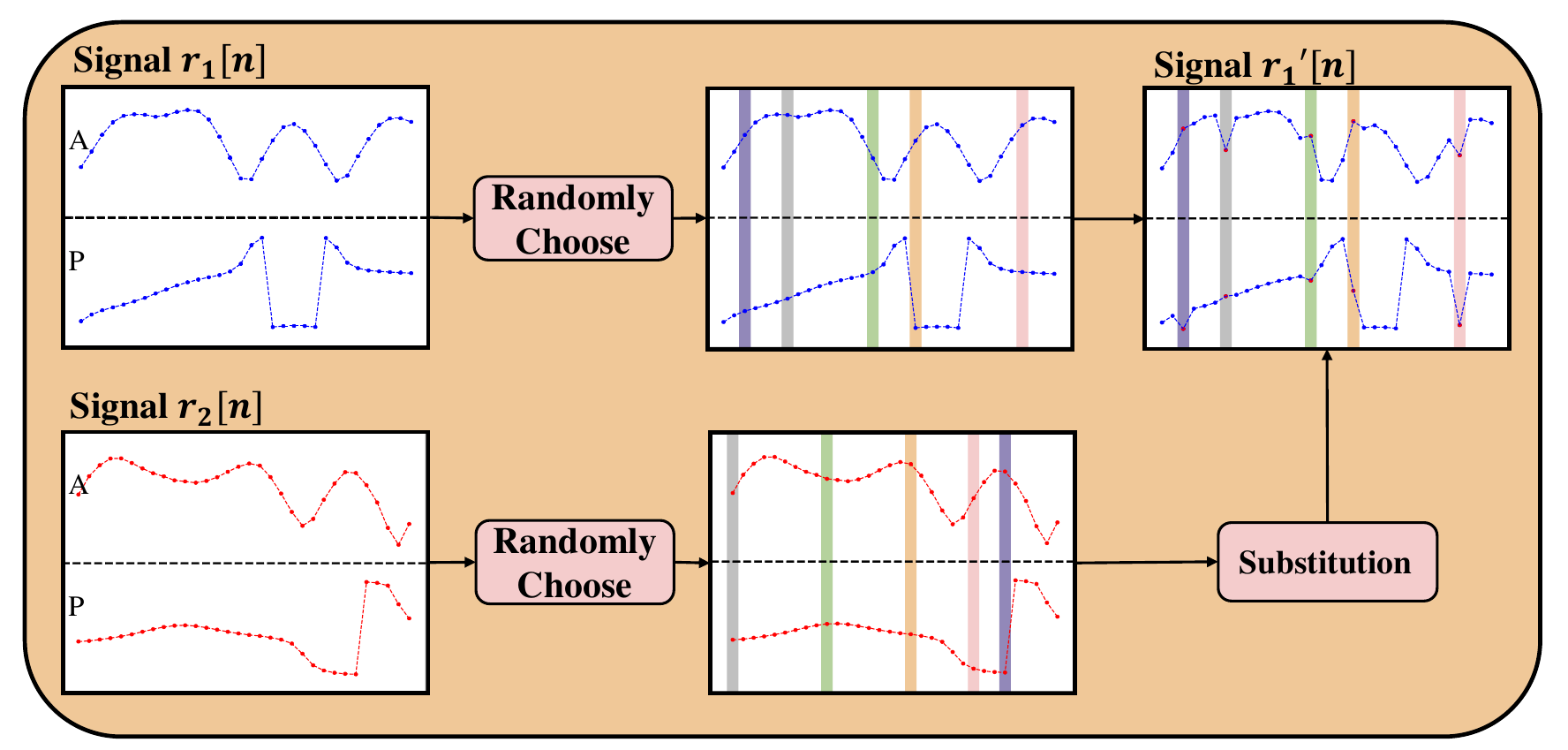}  }  
    \subfigure[Continuous Segment Substitution]{  \includegraphics[width=0.95\linewidth]{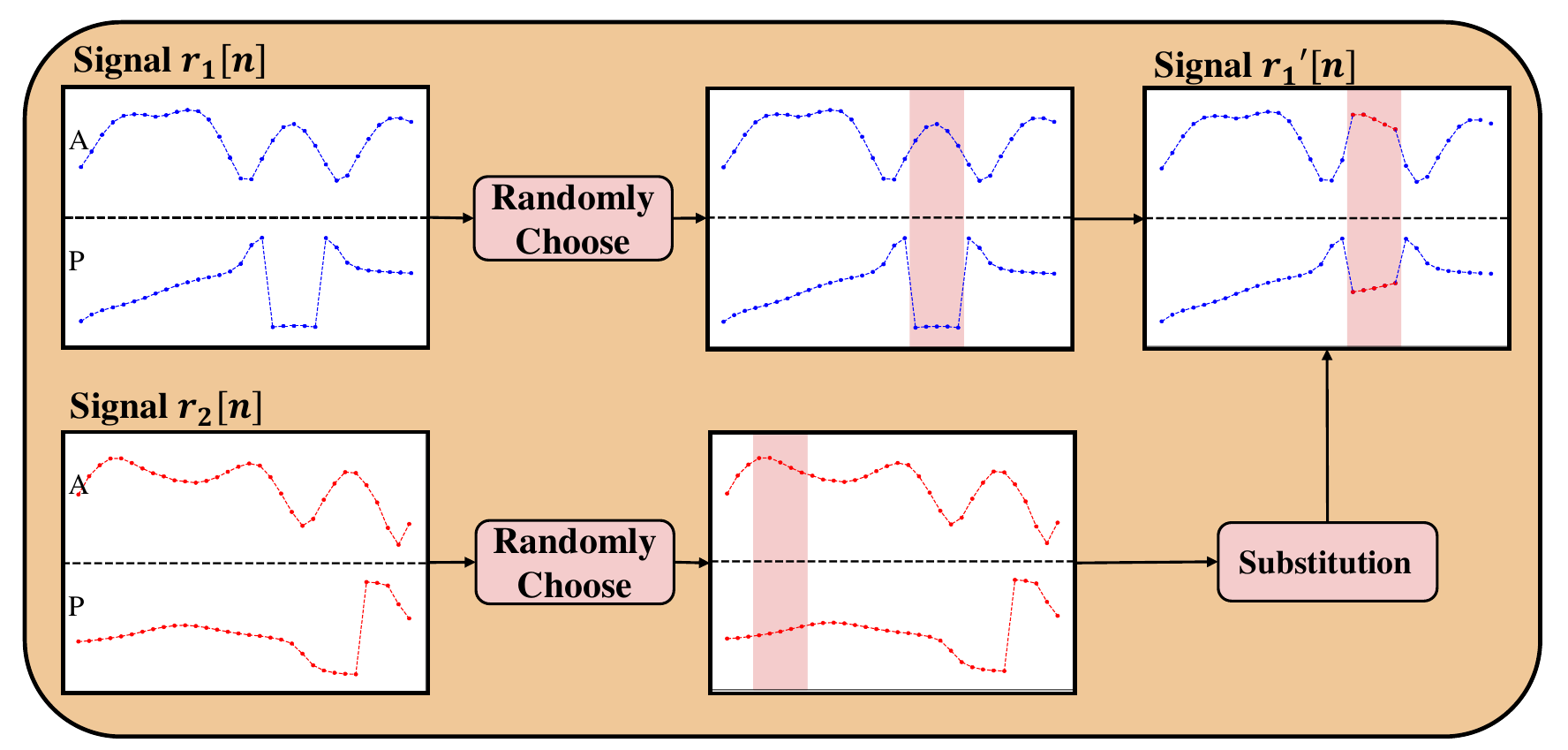}  }  
\caption{Principle of segment substitution strategies.}   
\label{randommix_principle}
\end{center}
\end{figure}
Building upon the aforementioned analysis, we propose two SS strategies.
In detail, for an original A/P signal with a length of $N$, we randomly choose samples of length $l$ and substitute them with other segments. 
Based on different ways of selecting and replacing sample points, SS strategies can be classified into the following two categories:
\subsubsection{Discrete Segment Substitution}
\label{dis}
As shown in Fig. \ref{randommix_principle} (a), we randomly choose $l$ discrete samples from signal $\boldsymbol{r}_1[n]$ for substitution, where the index set of these samples is denoted as $\mathcal{S}_1$. 
The selected sample points are delineated by distinct colored regions in the figure.
The segments used for Substitution are randomly chosen discrete samples with an index set $\mathcal{S}_2$, extracted from signal $\boldsymbol{r}_2[n]$ which has the same modulation type and a lower or equal SNR ratio with $\boldsymbol{r}_1[n]$.
These points are also annotated in the figure, corresponding to the regions of the same color in $r_1[n]$ for substitution.
Incorporating sample points with lower SNR ratios for substitution enhances the model's capability to handle noise and improve its robustness.
The entire process can be expressed as:
\begin{equation}
    \begin{split}
      \mathcal{S}_{1} = \{i_1, i_2, &..., i_k, ..., i_l\}, ~\text{where}~ 0\le i_k< N \\
     \mathcal{S}_{2} = \{j_1, j_2, &..., j_k, ..., j_l\}, ~\text{where}~ 0\le j_k< N \\
     & \boldsymbol{r}_1[\mathcal{S}_{1}] =  \boldsymbol{r}_2[\mathcal{S}_{2}]
     \\
    \end{split}
\end{equation}

\subsubsection{Continuous Segment Substitution}
As shown in Fig. \ref{randommix_principle} (b), We randomly choose continuous samples of length $l$ from $\boldsymbol{r}_1[n]$ and replace them with other segments, which are continuous samples extracted from signal $\boldsymbol{r}_2[n]$  with the same modulation type and a lower or equal SNR.
Both selected regions from the two signals are marked in the figure.
\begin{equation}
    \begin{split}
     \boldsymbol{r}_1[i:i+l] &=  \boldsymbol{r}_2[j:j+l],  \\
     ~\text{where}~ 0\le i&, j < N-l\\
    \end{split}
\end{equation}

Both of these approaches are advantageous in strengthening the properties of the modulation scheme, thus enabling the extraction of more generalized features.
From another perspective, SS strategies can generate new samples without altering the modulation scheme, thereby expanding the dataset. 
This is particularly important in few-shot scenarios. 
Such scenarios are frequently encountered, as obtaining non-homogeneous datasets is challenging in real-world environments due to the complexity of acquiring labeled signals.

\section{Experimental Results}
\subsection{Datasets and Setup}

\begin{table*}[!t]
  \footnotesize
\caption{Training Parameters for two Datasets}
\begin{center}
\label{Dataset}
\begin{tabular}{c | c | c }
    \toprule
\textbf{Parameters} & \textbf{RadioML2016.10a} & \textbf{RadioML2018.01a}\\
    \midrule
& & 32PSK, 16APSK, 32QAM, FM, GMSK, \\
& 8PSK, AM-DSB, AM-SSB, BPSK, & 32APSK, OQPSK, 8ASK, BPSK, AM-DSB-WC,\\
Modulation type &  CPFSK, GFSK, PAM4, QAM16, &  8PSK, AM-SSB-SC, 4ASK, 16PSK, OOK\\
& QAM64, QPSK, WBFM &   64APSK, 128QAM, 128APSK, AM-DSB-SC, 16QAM\\
& &   AM-SSB-WC, 64QAM, QPSK, 256QAM\\
    \midrule
SNR & -20:2:18 (dB) & -20:2:30 (dB)\\
    %\midrule
Total Sample Numbers & 220000 & 2555904\\
    %\midrule
Signal Length & 128 & 1024\\
    \midrule
Division & 6:2:2 & 6:2:2\\
    %\midrule
Batchsize & 128 & 512\\
    %\midrule
Loss Function & Cross Entropy & Cross Entropy\\
    %\midrule
Optimizer & AdamW & AdamW\\
    %\midrule
Initial Learning Rate & 0.001 & 0.001\\
    %\midrule
Learning Rate Scheduler& ReduceLROnPlateau & ReduceLROnPlateau\\
    %\midrule
Maximum Epochs & 150 & 150\\
\bottomrule
\end{tabular}
\end{center}
\end{table*}
We conduct experiments on our proposed algorithm using the widely adopted open-source datasets, RadioML2016.10a \cite{o2016radio} and RadioML2018.01a \cite{o2018over}. 
The signals in these datasets are synthetically generated using a signal generator and captured using GNU Radio. The transmission channel model incorporates Gaussian noise and multipath fading. The channel model considers various stochastic imperfections, such as sampling rate offset, center frequency offset, and fading, to simulate real-world environments.
The training parameters and dataset-related information are presented in Table \ref{Dataset}.
We divide the datasets for training, validation, and testing by a ratio of 6:2:2. We utilize AdamW \cite{loshchilov2017fixing} as the optimizer with an initial learning rate set to 0.001. We employ the ReduceLROnPlateau scheduler where the learning rate is reduced by a factor of 0.1 when the validation loss does not decrease for 10 epochs.

Below are the descriptions of the hyperparameters of the model. We trained two separate models on the two datasets, and their architectures slightly differ.
Due to the difference in signal lengths between the two datasets, we set the number of convolutional layers to $K=2$ for RadioML2016.10a and $K=4$ for RadioML2018.01a. 
This adjustment is made to reduce the sequence length of the feature embeddings, thereby minimizing model complexity and eliminating redundant information.
This exemplifies the expandability of the proposed backbone, as simply modifying the number of convolutional layers based on the signal length allows it to be applied to various datasets.
Each symbol has a sample size of $L_s = 8$ in both datasets and we set the convolutional kernel size $K_s = L_s/2$.
The number of convolutional output channels $d$ is set to 64. We have the feature embeddings $\boldsymbol{X}_e\in \mathbb{R}^{32\times 64} $ for RadioML2016.10a and $\boldsymbol{X}_e\in \mathbb{R}^{64\times 64} $ for RadioML2018.01a.

The rest of the hyperparameters remain the same for both datasets. 
We set the reduction ratio of the SE block to $r=4$. In the transformer, the layer number is $M_t=2$ with attention head number $h=8$, and the dimension of the feed-forward layer is $d_{ffn} = 2d$. 
The number of LSTM layers is set to $M_l=4$, and the hidden dimension is $d_l = d$.

We compare our proposed TLDNN with several well-known SOTA methods as baselines, including ResNet \cite{o2018over}, LSTM2 \cite{rajendran2018deep}, LSTM-DAE \cite{ke2021real}, MCLDNN \cite{xu2020spatiotemporal}, MCformer \cite{9685815}, and FEA-T \cite{9915584}. 
Among them, MCformer and FEA-T are both transformer-based models. Due to its high complexity, MCformer is not evaluated on the RadioML2018.01a dataset.
All experiments are conducted using the PyTorch framework and implemented on an NVIDIA GeForce GTX 3090 GPU.

\subsection{Performance Comparison With Other Methods}
\begin{figure*}[!t]
\begin{center}  
    \subfigure[Performance on RadioML2016.10a]{  \includegraphics[width=0.45\linewidth]{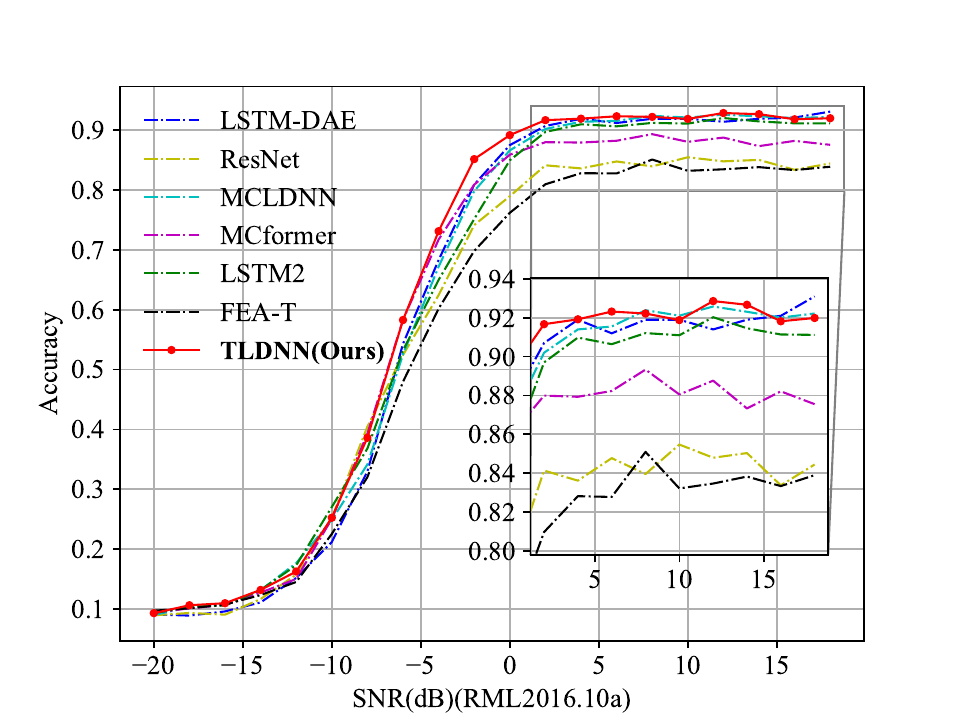}  }  
    \subfigure[Performance on RadioML2018.01a]{  \includegraphics[width=0.45\linewidth]{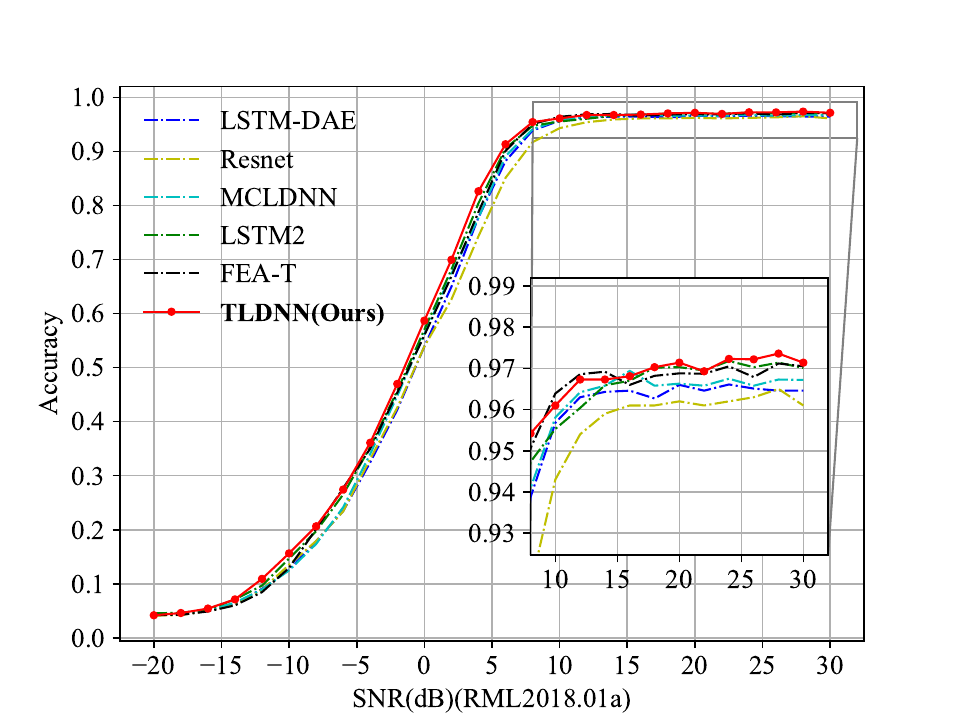}  }  
\caption{Recognition performance with other deep learning methods.}   
\label{sota_comparison}
\end{center}
\end{figure*}

In this section, we compare our proposed TLDNN with other state-of-the-art deep learning methods. 
Additionally, we present the results obtained after applying the SS strategy.
The inputs for ResNet, MCLDNN, and MCformer are the raw I/Q signals, while the inputs for LSTM2, LSTM-DAE, and our proposed TLDNN are the A/P signals. 
In particular, FEA-T inputs are embedding tokens generated by sliding window framing of I/Q signals.
The curve illustrating the variation of recognition accuracy with respect to SNR is depicted in  Fig. \ref{sota_comparison}.
In addition, the complexity and other accuracy metrics for all benchmarks are presented in Tabl \ref{sota_tabel}.

\begin{table*}
\footnotesize
\centering
\caption{Performance comparison of all methods}
\label{sota_tabel}
\begin{threeparttable}
\begin{tabular}{ c | cc |ccc |cc | cc c }
\toprule
\multirow{3}{*}{\textbf{Benchmarks}} &  \multicolumn{5}{c|}{\textbf{RadioML2016.10a}} &  \multicolumn{5}{c}{\textbf{RadioML2018.01a}}\\
\cmidrule(lr){2-11}

~ &  \multirow{2}{*}{Params} & \multirow{2}{*}{FLOPs} & \multicolumn{3}{c|}{Accuracy(\%)} & \multirow{2}{*}{Params} & \multirow{2}{*}{FLOPs} & \multicolumn{3}{c}{Accuracy(\%)}\\
%\cline{4-6}
%\cline{9-11}
~ &  ~ & ~ & Maximum & LowSNR\tnote{1} & Average &  ~ & ~ & Maximum & LowSNR\tnote{2} & Average\\
\midrule
ResNet\cite{o2018over} &  85.52K  & 3.24M & 85.5 & 52.4 & 57.32 &  164.18K & 25.94M & 96.5 & 42.6 & 60.91\\

LSTM-DAE\cite{ke2021real} &  14.99K  & 1.75M & 93.1 & 54.3 & 61.42 &  15.21K & 13.96M & 96.6 & 42.2 & 61.32\\

MCformer\cite{9685815} &  73.16K  & 7.76M & 89.3 & 58.3 & 60.54 & 74.78K & 324.26M & NA\tnote{3} & NA &NA \\

FEA-T\cite{9915584} &  269.13K  & 2.19M & 85.1 & 48.0 & 55.74 &  269.98K & 19.66M & 97.1 & 45.2 & 62.37\\

MCLDNN\cite{xu2020spatiotemporal} &  368.65K  & 41.88M & 92.6 & 52.5 & 61.52 &  370.33K & 343.16M & 96.7 & 44.8 & 61.92\\
 
LSTM2\cite{rajendran2018deep} &  199.05K  & 25.56M & 92.0 & 53.2 & 61.02 &  200.73K  & 204.48M & 97.2 & 45.7 & 62.52\\
\midrule
\textbf{TLDNN} &  243.34K  & 7.89M & 92.9 & 58.3 & 62.83 &  276.66K  & 22.89M & \textbf{97.4} & 47.0 & 63.32\\
\textbf{TLDNN (\textit{+ SS})} &  243.34K  & 7.89M & \textbf{93.4} & \textbf{59.6} &  \textbf{63.35} &  276.66K  & 22.89M & \textbf{97.4} & \textbf{47.4} & \textbf{63.42}\\
\midrule
 \textit{Improvement to Existing Best}\tnote{4} & \textit{-34.0\%}& \textit{-81.2\%} &\textit{+0.3\%}&\textit{+11.0\%}&\textit{+2.1\%}& \textit{+37.8\%}& \textit{-88.8\%} &\textit{+0.2\%}&\textit{+2.9\%} &\textit{+1.3\%}\\
\bottomrule
\end{tabular}
\begin{tablenotes}    % 添加命令
        \footnotesize               % 添加命令
        \item[1] We utilize the accuracy at -6dB as low SNR conditions in RadioML2016.10a. The subsequent experiments employ the same evaluation metrics.
        \item[2] We utilize the accuracy at -2dB as low SNR conditions in RadioML2018.01a. The subsequent experiments employ the same evaluation metrics.
        \item[3] MCformer necessitates an extensive allocation of memory for training, rendering it infeasible. The original paper lacks a report on RadioML2018.01a.
        \item[4] We compare our original TLDNN with the best-performing baseline methods, which are MCLDNN in RadioML2016.10a and LSTM2 in RadioML2018.01a.
      \end{tablenotes}            % 添加命令

\end{threeparttable}
\end{table*}

As shown in Tab. \ref{sota_tabel} and Fig. \ref{sota_comparison} (a), our proposed TLDNN model outperforms existing SOTA methods on Radio2016.10a across all SNR levels and achieves the highest average accuracy of 62.83\% (+2.1\%). 
Particularly, TLDNN demonstrates remarkable performance in the mid-to-low SNR range, exhibiting similar performance to MCformer and significantly outperforming other methods.
TLDNN model achieves an accuracy of 58.3\% at -6dB as representative of low SNR.
In high SNR scenarios, TLDNN achieves the highest accuracy of 92.9\% at 12dB, whereas MCformer lags with a maximum accuracy of 89.3\%.
This indicates that TLDNN possesses a unique advantage over other models in dealing with noise, demonstrating superior performance in low SNR environments while maintaining generalizability across all noise levels.

The results presented in Tab. \ref{sota_tabel} and Fig. \ref{sota_comparison} (b) for the Radio2018.01a dataset indicate that TLDNN outperforms other baseline methods across all SNR levels, achieving the highest average accuracy of 63.32\% (+1.3\%). 
TLDNN continues to exhibit significantly superior performance in the mid-to-low SNR range, achieving the highest accuracy of 47.3\% at -2dB. 
In high SNR conditions, TLDNN achieves the highest maximum accuracy 97.4\% at 28dB. 
The above results demonstrate that our TLDNN structure, serving as the foundational backbone, achieves remarkable SOTA performance while maintaining excellent generalization ability across diverse datasets.
After incorporating the SS, TLDNN exhibits a noticeable improvement, which will be detailly validated in Sec. \ref{exp:randommixing}.

In terms of complexity analysis, we compare the number of parameters and floating-point operations per second (FLOPS) among different models. The complexity metrics of the model are computed using the THOP (PyTorch-OpCounter) library.
TLDNN has a higher parameter count compared to the less effective models, ResNet and LSTM-DAE, but it is comparable to other models. This suggests that our model does not require larger memory space. 
However, our method exhibits significantly improved performance compared to ResNet and LSTM.
In terms of computational speed, TLDNN outperforms the existing best models, MCLDNN and LSTM2, by achieving a substantial reduction of 80\% to 90\% in FLOPs.

\begin{figure*}[!t]
\centering
\includegraphics[width=0.71\linewidth]{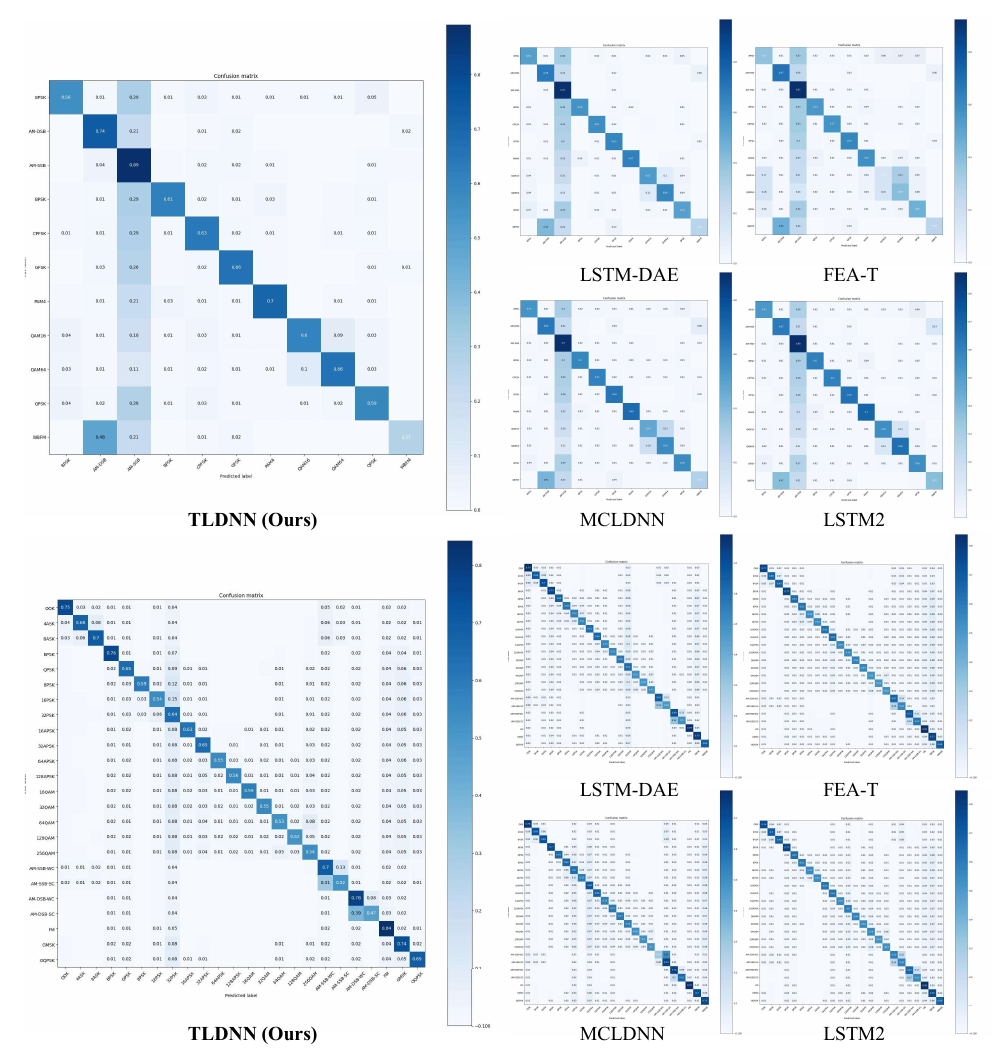}
\caption{Confusion matrices of TLDNN (left), LSTM-DAE, FEA-T, MCLDNN, and LSTM2 (right) models, on RadioML2016.10a (top) and
RadioML2018.01a (bottom)}
\label{confusion}
\end{figure*}
To provide a more comprehensive analysis, we visualize the confusion matrices of LSTM-DAE, FEA-T, MCLDNN, LSTM2, and TLDNN on the test sets, as shown in Fig. \ref{confusion}. 
In RadioML2016.10a, FEA-T, MCLDNN, and LSTM2 models tend to confuse the QAM16 and QAM64 modulation schemes, while TLDNN performs best in classifying these two types. All models tend to misclassify WBFM as AM-SSB, indicating that these two analog modulation schemes are very similar and difficult to distinguish.
In RadioML2018.01a, both AM-SSB-WC and AM-SSB-SC schemes belong to single-side band amplitude modulation and are prone to confusion. This observation aligns with our understanding, as these two schemes differ only in whether the carrier is suppressed.
Similarly, AM-DSB-WC and AM-DSB-SC, which belong to double-side band amplitude modulation, can easily be confused. FEA-T and TLDNN demonstrate the best performance in recognizing these four modulation schemes, indicating that the transformer structure possesses unique advantages in identifying whether a signal contains a carrier component.

\subsection{Effectiveness of Individual Components}
\begin{figure*}[htbp]
\begin{center}  
    \subfigure[Performance on RadioML2016.10a]{  \includegraphics[width=0.45\linewidth]{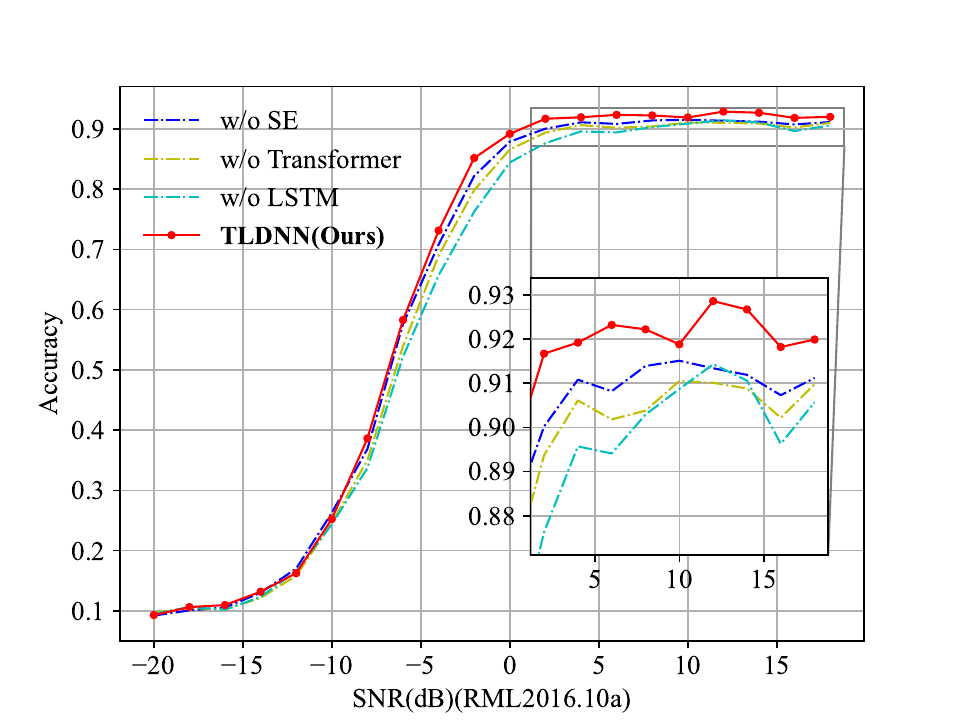}  }  
    \subfigure[Performance on RadioML2018.01a]{  \includegraphics[width=0.45\linewidth]{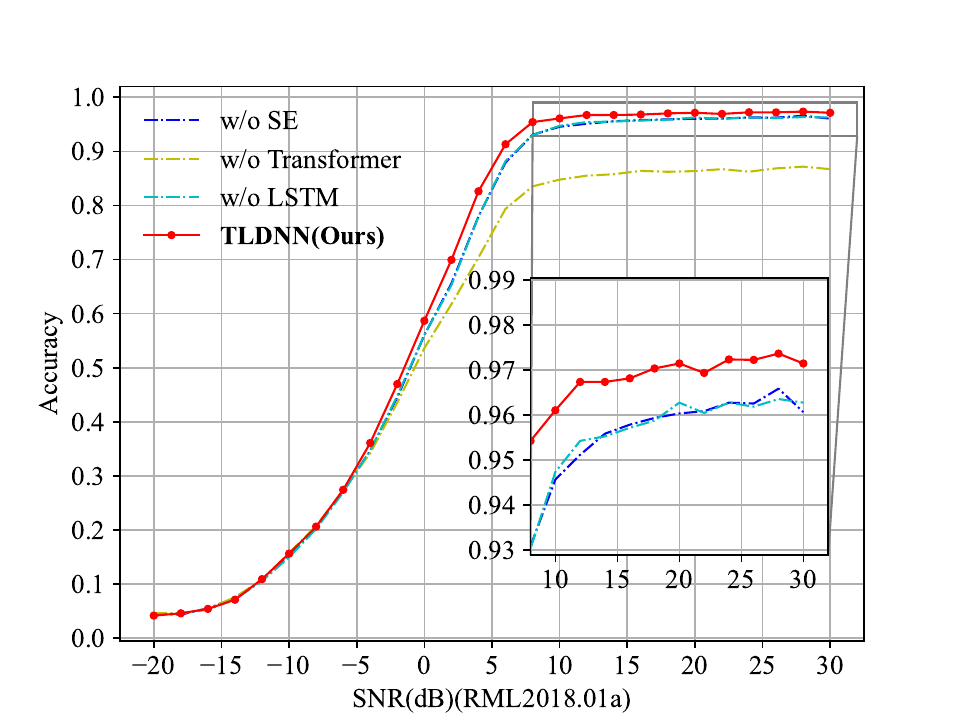}  }  
\caption{Ablation study of the TLDNN architecture: compared with its variants with different components removed.}   
\label{Ablation}
\end{center}
\end{figure*}
We conduct an ablation study to inspect the contribution of each component in the overall TLDNN architecture. 
Due to the necessity of including the convolutional layers to maintain the dimensionality of the feature embeddings, we selectively remove other components of TLDNN to evaluate the effectiveness of each module.
We compare the experimental results of the complete TLDNN with three variations: removing the SE block, removing the transformer module, and removing the LSTM module. The results are shown in Fig. \ref{Ablation}.

The complete model performs best on both datasets and removing each module leads to a decrease. This demonstrates the complementary advantages of each module, and their combination forms a superior model.
We observe that removing the SE block has a relatively small impact on the performance. This is consistent with our expectations because the SE block primarily performs channel-wise weighting on the outputs of the convolutional layer, without incorporating additional feature information from input signals. 

In RadioML2016.10a, the model without LSTM shows the worst performance, while it performs similarly to the model without SE block in RadioML2018.01a. 
We believe that in the case of shorter signal lengths, the presence of LSTM is crucial for capturing temporal dependencies and sequence information.
In RadioML2018.01a, removing the transformer module results in significant performance degradation, surpassing other cases. We attribute this to the longer length of the signals, which provides more information for the transformer to extract global features effectively. 
\subsection{Different Internal Structure of transformer}
\iffalse
\begin{figure*}[htbp]
\begin{center}  
    \subfigure[Performance on RadioML2016.10a]{  \includegraphics[width=0.45\linewidth]{figure/init_2016.pdf}  }  
    \subfigure[Performance on RadioML2018.01a]{  \includegraphics[width=0.45\linewidth]{figure/init_2018.pdf}  }  
\caption{Performance of different internal structures of the transformer (The abbreviation "TH" is employed to represent the term "talking-heads attention").}   
\label{Init}
\end{center}
\end{figure*}
\fi

\begin{table*}
\centering
\footnotesize
\caption{Performance of different internal structures of the transformer}
\label{Init}
\begin{tabular}{ cc | ccc |ccc}
\toprule
\multicolumn{2}{c|}{\textbf{Modification}} &  \multicolumn{3}{c|}{\textbf{RadioML2016.10a Accuracy(\%)}} &  \multicolumn{3}{c}{\textbf{RadioML2018.01a Accuracy(\%)}}\\
Talking-heads & ReGLU  & Maximum & LowSNR & Average &  Maximum & LowSNR & Average\\
\midrule
\XSolidBrush&\XSolidBrush& 91.3  & 56.0 & 61.64 &  96.5 & 44.3 & 61.74\\

\Checkmark& \XSolidBrush & 92.0 & 57.6 & 62.08 &  96.8 &  45.3 & 62.26\\

\XSolidBrush&\Checkmark&  92.6 & 57.6 &  62.51 & 97.0  & 47.0 & 62.84\\
\midrule
\Checkmark & \Checkmark & \textbf{92.9} & \textbf{58.3} & \textbf{62.83} & \textbf{97.4} & \textbf{47.0} & \textbf{63.32}\\
\bottomrule
\end{tabular}
\end{table*}

This section aims to validate the effectiveness of the internal structure adjustments made to the transformer module in our proposed TLDNN.
The position-wise FFN in the original transformer consists of a simple multi-layer perceptron (MLP) network, which is replaced with ReGLU layers in our TLDNN. 
We introduce the concept of talking-heads, which establishes connections between different heads to enhance the representational capacity of the model.

We conduct experiments using four different configurations in TLDNN: with and without the application of each modification.
When the linear mapping dimensions are the same, the number of parameters in ReGLU is approximately 1.5 times that of MLP. 
To ensure a fair comparison under similar complexity conditions, we control the parameter sizes of both models to be roughly equivalent. 
The application of the talking-heads mechanism only leads to a marginal increase in the size of parameters, which can be considered negligible. The results are shown in Fig. \ref{Init}.

By observing the experimental results, we can see that the original transformer model performs the worst. Introducing either the ReGLU or the talking-heads mechanism improves the accuracy of recognition. 
This result indicates that the talking-heads mechanism facilitates the exchange of information between different heads.
Meanwhile, ReGLU utilizes gate units to extract more expressive features within each embedding. 
They both contribute to promoting information transfer within each token embedding. The best performance is achieved when these two improvements are combined, demonstrating the effectiveness of our proposed improvements without increasing complexity. 

\subsection{Adjusting the Depth of the Model}
\begin{table*}
\centering
\footnotesize
\caption{Performance of different transformer layer number}
\label{transformerlayer}
\begin{tabular}{ c | cc|ccc | cc | ccc}
\toprule
\multirow{3}{*}{\makecell{\textbf{Transformer}\\ \textbf{Layer Number $M_t$}}} &  \multicolumn{5}{c|}{\textbf{RadioML2016.10a}} &  \multicolumn{5}{c}{\textbf{RadioML2018.01a}}\\
\cmidrule(lr){2-11}
~ &  \multirow{2}{*}{Params} & \multirow{2}{*}{FLOPs} & \multicolumn{3}{c|}{Accuracy(\%)} & \multirow{2}{*}{Params} & \multirow{2}{*}{FLOPs} & \multicolumn{3}{c}{Accuracy(\%)}\\
%\cline{4-6}
%\cline{9-11}
~ &  ~ & ~ & Maximum & LowSNR & Average &  ~ & ~ & Maximum & LowSNR & Average\\
\midrule
2 &  243.34K & 7.89M & 92.9 & 58.3 & 62.83 &  276.66K  & 22.89M & 97.4 & 47.0 & 63.32\\
%\hline
4 &  327.69K & 10.89M & 92.5 & 57.9 & 62.63 & 361.02K & 29.50M & \textbf{97.4} & 47.7 & \textbf{63.53}\\
%\hline
6 & 412.04K & 13.89M & \textbf{93.1} & \textbf{59.4} & \textbf{63.13} &  445.37K & 36.12M & 97.3 & 47.4 & 63.42\\
%\hline 
8 &  496.40K &16.89M & 91.1 & 55.7 & 61.40 &  529.72K & 42.74M & \textbf{97.4} & \textbf{48.0} & 63.45 \\
\bottomrule
\end{tabular}
\end{table*}

\begin{table*}
\footnotesize
\centering
\caption{Performance of different LSTM layer number}
\label{LSTMlayer}
\begin{tabular}{ c | cc|ccc | cc | ccc}
\toprule
\multirow{3}{*}{\makecell{\textbf{LSTM}\\ \textbf{Layer Number $M_l$}}} &  \multicolumn{5}{c|}{\textbf{RadioML2016.10a}} &  \multicolumn{5}{c}{\textbf{RadioML2018.01a}}\\
\cmidrule(lr){2-11}
~ &  \multirow{2}{*}{Params} & \multirow{2}{*}{FLOPs} & \multicolumn{3}{c|}{Accuracy(\%)} & \multirow{2}{*}{Params} & \multirow{2}{*}{FLOPs} & \multicolumn{3}{c}{Accuracy(\%)}\\
%\cline{4-6}
%\cline{9-11}
~ &  ~ & ~ & Maximum & LowSNR & Average &  ~ & ~ & Maximum & LowSNR & Average\\
\midrule
2 &  176.78K &5.73M & 91.5 & 56.2 & 61.50 &  210.10K & 18.56M & 97.3 & 46.6 & 63.11\\
%\hline
4 &  243.34K & 7.89M & \textbf{92.9} & \textbf{58.3} & \textbf{62.83} &  276.66K  & 22.89M & 97.4 & 47.0 & 63.32\\
%\hline
6 &  309.90K & 10.05M & 92.4 & 58.0 & 62.40 &  343.22K & 27.21M & 97.3 & \textbf{47.3} & 63.22\\
%\hline 
8 &  376.46K & 12.22M & 91.3 & 57.0 & 61.96 &  409.78K & 31.54M & \textbf{97.5} & 47.1 & \textbf{63.34}\\
\bottomrule 
\end{tabular}
\end{table*}
In this section, we focus on investigating the impact of adjusting the depth of the model architecture. We aim to understand how changes in the model's depth affect its ability to learn and represent the underlying patterns in the data.

First, we fix the number of LSTM layers at $M_l=4$ and adjust the depth of the transformer layers $M_t$ from 2 to 8 in increments of 2. The results are shown in Tab. \ref{transformerlayer}.
On RadioML2016.10a, increasing the depth can lead to performance improvements. The highest average accuracy of 63.13\% is achieved when $M_t=6$. 
However, when the model is deeper, severe overfitting occurs, resulting in a significant accuracy drop. 
On RadioML2018.01a, the optimal results are achieved when $M_t=4$. 
As the depth increases, the performance reaches a plateau.
Considering the increased computational cost of adding layers, we believe that selecting a depth of 2 strikes the optimal balance between the two factors.

Next, we fix the number of transformer layers at $M_t=2$ and adjust the depth of the LSTM layers $M_l = 2,4,6,8$, respectively. The results are presented in Tab. \ref{LSTMlayer}. 
On Radio2016.10a, we achieve the best performance when the $M_l=4$. As the number of layers increases, the performance gradually deteriorates, indicating a severe issue of overfitting. It becomes evident that deeper structures fail to extract additional meaningful information, leading to diminishing returns.
On RadioML2018.01a, when $M_l$ is set to 4, it achieves near-optimal performance. Expanding the depth has minimal impact on the performance.
Therefore, we assert that a 4-layer LSTM architecture is the most suitable choice, as it achieves optimal performance within the constraints of computational cost.

The divergent performances observed on these two datasets can be attributed to the difference in signal lengths. 
In RadioML2016.10a, the signals are relatively short, consisting of only 128 samples. A shallow network is sufficient for effective feature extraction while deeper networks are more prone to overfitting in such scenarios.
We believe that the backbone can be appropriately adjusted based on the length of the signals, with deeper networks being suitable for longer signals.

\subsection{Effectiveness of Segment Substitution}
\label{exp:randommixing}
In this paper, we propose two data augmentation strategies known as segment substitution to overcome the influence of modulation-independent features such as RF fingerprint characteristics, channel features, and pilot regularities, thereby improving the generalization performance. 
In practice, due to the sake of reducing memory usage, we create a subset for choosing substitution segments by selecting several samples from each modulation scheme.
Experiments show that this approach provides sufficient randomness for augmentation.

\begin{figure*}[!t]
\begin{center}  
    \subfigure[Performance on RadioML2016.10a]{  \includegraphics[width=0.45\linewidth]{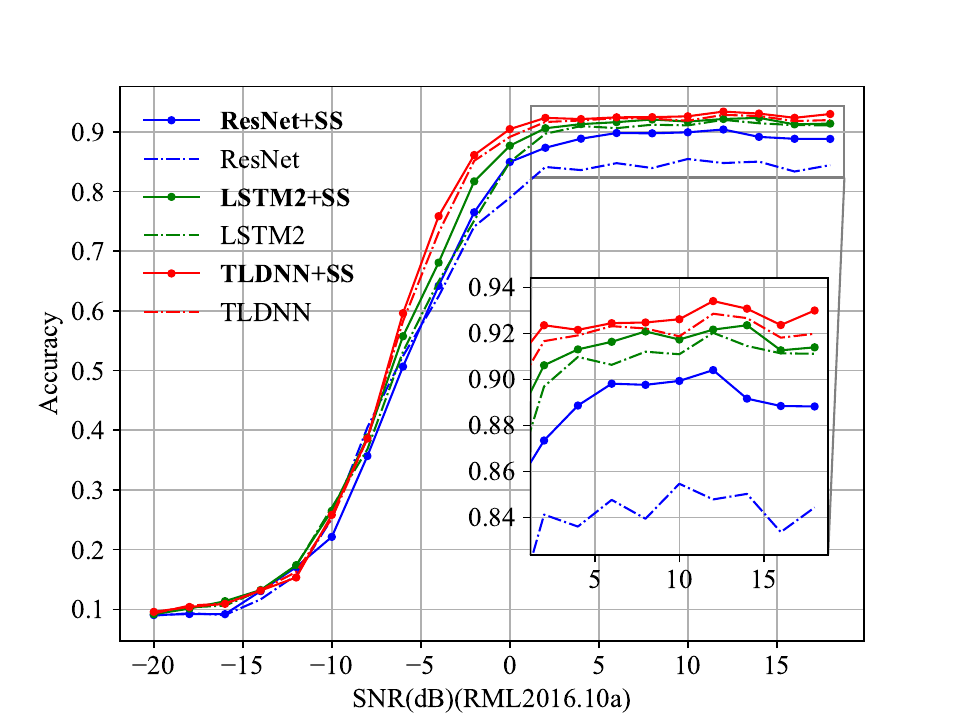}  }  
    \subfigure[Performance on RadioML2018.01a]{  \includegraphics[width=0.45\linewidth]{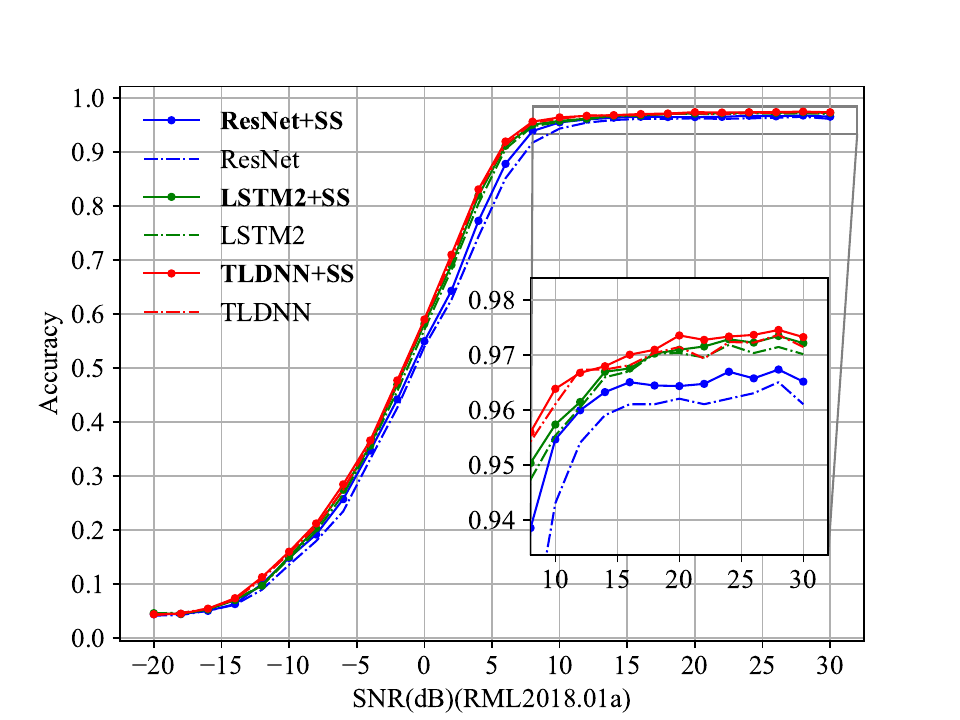}  }  
\caption{Performance comparison with models with and without \textit{Discrete Segment Substitution}.}   
\label{randommixing_fig}
\end{center}
\end{figure*}

We adjust the substitution ratio $r=\frac{l}{N}$ based on the signal length $N$ in each dataset.
A low ratio may not fully exploit the advantages of augmentation, while a high ratio may introduce excessive contamination of the signal. 
Based on our experimentation, we set the optimal ratio as $r=\frac{1}{16}$ for RadioML2016.10a and $r=\frac{1}{64}$ for RadioML2018.01a.
We apply our strategies to ResNet, LSTM2, and our TLDNN to validate the effectiveness of segment substitution across different methods. 
We compare our methods with the commonly used noise addition method in RF signal recognition,  which can be performed by adding Gaussian noise to the received signals with a standard deviation $\sigma=0.0001$
\cite{huang2019data,chae2022rethinking}.
The experimental results are shown in Fig. \ref{randommixing_fig} and Tab. \ref{strategies}.

\begin{table*}
\centering
\footnotesize
\caption{Performance of different data augmentation strategies}
\label{strategies}
\begin{threeparttable}
\begin{tabular}{ c| c | ccc | ccc}
\toprule
\multirow{2}{*}{\textbf{Benchmarks}} & \multirow{2}{*}{\makecell{\textbf{Augmentation}\\ \textbf{Strategies}}} &  \multicolumn{3}{c|}{\textbf{RadioML2016.10a Accuracy(\%)}} &  \multicolumn{3}{c}{\textbf{RadioML2018.01a Accuracy(\%)}}\\
%\cmidrule{lr}{3-8}
~  & ~  & Maximum & LowSNR & Average &  Maximum & LowSNR & Average\\
%\cmidrule{lr}{1-8}
\midrule
\multirow{5}{*}{ResNet\cite{o2018over}}  & NA & 85.5 & 52.4 & 57.32 & 96.5 & 42.6 & 60.91\\
~  & \textit{Noise Addition} &  85.7 & 49.6 & 56.96 &  \textbf{96.8} & 43.2 & 61.46\\
~  & \textit{Discrete Segment Substitution} & \textbf{90.4} & \textbf{50.7}  & \textbf{59.84}  & 96.4 & \textbf{43.9} &  \textbf{61.50}\\
~  & \textit{Continuous Segment Substitution} & 89.7  & 48.9 & 59.24 &  96.3 & 43.5 & 61.32\\
\midrule

\multirow{5}{*}{LSTM2\cite{rajendran2018deep}}  & NA & 92.0 & 53.2 & 61.02 & 97.2 & 45.7 & 62.52\\
~  & \textit{Noise Addition} &  92.3 & 52.9 & 61.38 &  97.1 & \textbf{46.9} & 62.62\\
~  & \textit{Discrete Segment Substitution} & 92.4 & \textbf{55.9} &  62.19 & \textbf{97.2} & \textbf{46.9} & \textbf{62.89} \\
~  & \textit{Continuous Segment Substitution} &  \textbf{93.0} & \textbf{55.9} & \textbf{62.40} &  97.1  & 46.5 & 62.81 \\
\midrule

\multirow{5}{*}{\textbf{TLDNN(Ours)}}  & NA & 92.9 & 58.3 & 62.83 & 97.4 & 47.0 & 63.32\\
~  & \textit{Noise Addition} & 93.3  & 57.7 & 62.92 &   97.2 & 46.5 & 63.04 \\
~  & \textit{Discrete Segment Substitution} & \textbf{93.4} & \textbf{59.6} &  \textbf{63.35} & \textbf{97.4} & \textbf{47.6} & \textbf{63.42} \\
~  & \textit{Continuous Segment Substitution} & 92.7  & 57.8 & 62.80 &  97.3 & 46.4 & 63.14\\
\bottomrule
\end{tabular}

\end{threeparttable}
\end{table*}

It can be observed that after applying data augmentation, the performance of all the methods improves. 
The increase on RadioML2016.10a is significantly larger than the increase on RadioML2018.01a, which aligns with our expectations.
Due to the substantial size of the RadioML2018.01a dataset, overfitting is not prominent, and the impact of data augmentation on enriching dataset diversity is limited.
Another phenomenon is that poorly performing models, such as ResNet, show more significant improvement compared to well-performing models. Data augmentation enhances the model's ability to capture more universal features while improving its robustness against noise and other inherent interferences, which is particularly effective for models with poor performance.

Our segment substitution methods outperform noise addition significantly across different augmentation techniques.
Noise addition only considers the model's ability to handle noise, while our method, in addition to countering noise, disrupts time-domain relationships by reorganizing them. 
This disruption affects features such as pilot regularities and RF fingerprints that are sensitive to time relationships, allowing the model to focus more on modulation-related features.
\textit{Discrete Segment Substitution} outperforms \textit{Continuous Segment Substitution}, which aligns with our analysis of human recognition principles. The ability to determine modulation schemes based on global constellation diagrams is more precise and the discrete substitution approach is more conducive to extracting these global features.
In conclusion, our SS strategies have been validated for their effectiveness in improving model generalization. 
They outperform the noise addition method and exhibit robustness across various benchmarks and datasets.

\subsection{Segment Substitution in Few-Shot Learning}
\begin{figure*}[!t]
\begin{center}  
    \subfigure[Performance on RadioML2016.10a]{  \includegraphics[width=0.45\linewidth]{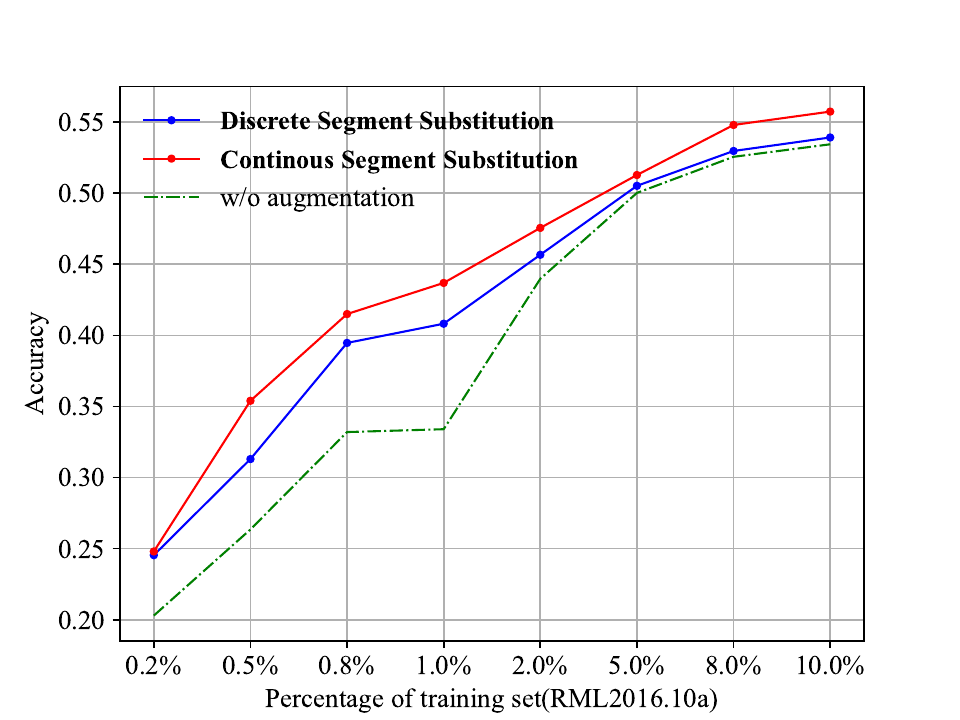}  }  
    \subfigure[Performance on RadioML2018.01a]{  \includegraphics[width=0.45\linewidth]{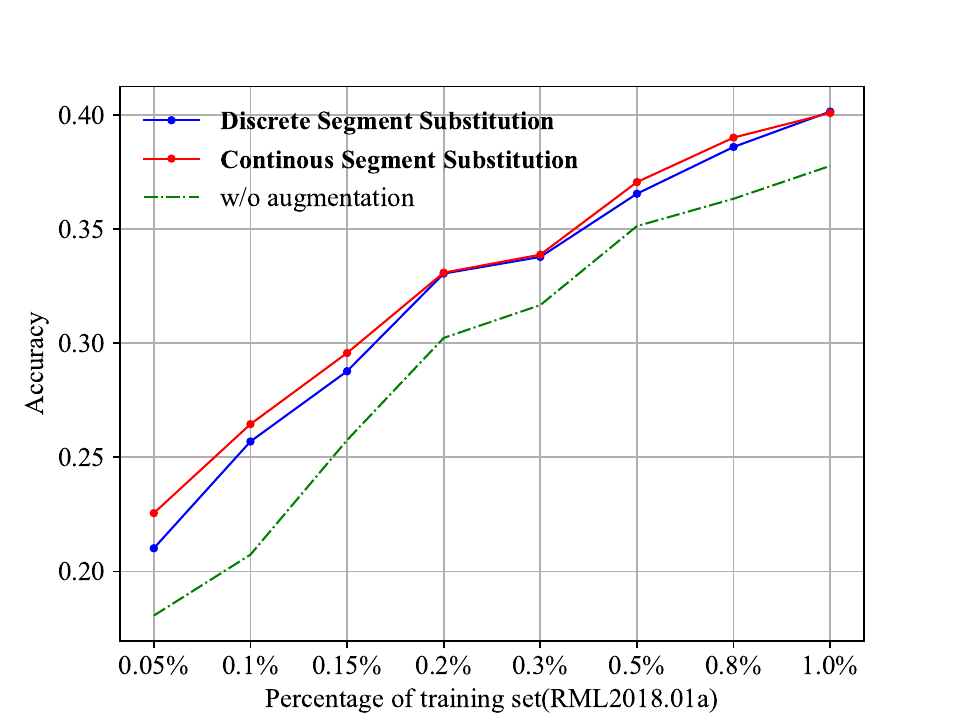}  }  
\caption{Performance comparison with and without segment substitution in few-shot learning.}   
\label{randommix_fewshot}
\end{center}
\end{figure*}
Collecting and annotating RF signals is a more intricate task compared to images and audio. This is due to the requirement for more specialized equipment, technical personnel, and a heightened emphasis on non-interference environmental conditions.
Constrained by these requirements, acquiring labeled RF signals becomes challenging. 
In practice, it is common to encounter AMR scenarios using only a small number of labeled samples, which is known as few-shot learning.
Few-shot learning is an exceptionally challenging task as it is constrained by the limited availability of modulation features for reference and induction.
Furthermore, few-shot learning presents another challenge specifically for the AMR task. 
Due to the high costs of changing the collection conditions, the samples are often collected in the same or similar environments.
This implies that they are heavily influenced by channel characteristics and RF fingerprint features to a significant extent. 
In scenarios with small samples, these influences become more pronounced as modulation features that are difficult to induce are more susceptible to being overwhelmed.

Our proposed segment substitution strategies allow for dataset expansion by reorganizing existing samples. 
Simultaneously, it emphasizes the features related to modulation schemes and mitigates the influence of environmental factors. 
We believe that our method addresses the two aforementioned challenges faced in few-shot learning, significantly enhancing the performance in such scenarios.
We validate the performance of TLDNN with and without segment substitution on RadioML2016.10a and RadioML2018.01a.
Based on the varying sizes of the dataset, we adjust the proportion of the training set $\eta$.
In the least case, we select only two samples per modulation category at each SNR level for training.
The substitution ratio for segment substitution is set to $r=1/8$ for both datasets.
The final results are depicted in Fig. \ref{randommix_fewshot}.

It can be observed that after applying the SS strategies, the performance of the model significantly improves on both datasets, resulting in a 5\%-10\% increase in accuracy.
This demonstrates that SS strategies effectively enhance the performance in few-shot learning scenarios, particularly when the proportion $\eta$ is smaller.
Even when $\eta$ increases, it still leads to a 3\% improvement in accuracy.
The larger improvement on RadioML2016.10a than RadioML2018.01a can be attributed to its smaller dataset size and shorter sample lengths, making the impact of the data augmentation strategy more pronounced in enhancing the diversity of the dataset.
\textit{Discrete Segment Substitution} outperforms \textit{Continuous Segment Substitution}, which aligns with the conclusion in Sec. \ref{exp:randommixing} that the discrete strategy better adheres to the principles of global recognition in constellation diagrams.
The above results suggest the generality of segment substitution in few-shot learning, enabling the model to learn generalized modulation patterns.

\section{Conclusion}
In this paper, we analyzed how human experts used constellation diagrams for modulation identification and discovered the key to AMR lies in extracting global structural features. 
Based on these observations, We introduced a hybrid network architecture called TLDNN, which combines the transformer to extract global correlations and LSTM to capture temporal dependencies. 
We incorporated the talking-heads attention and ReGLU FFN into the transformer to facilitate better interaction within token embedding.
Our proposed TLDNN achieved SOTA performance on the widely-used RadioML2016.10a and RadioML2018.01a datasets, especially in low SNR scenarios.
Our approach significantly reduces the complexity by 80\%-90\% compared to SOTA methods.
To mitigate the inherent impact of RF fingerprint and channel characteristics on generalization, we proposed data augmentation strategies known as segment substitution.
Our strategies effectively expand the dataset and enhance the model's generalization capability, particularly in few-shot scenarios that are commonly encountered in real-world environments.
We hope this work will inspire future research on robust global feature extraction in AMR.

\bibliographystyle{IEEEtran}
\bibliography{ref.bib}

\vfill

\end{document}